\pdfoutput=1
\documentclass[PAPER, cernpreprint, atlasdraft=false, UKenglish, texmf, orcidlogo]{atlasdoc}
\usepackage[subcaption]{atlaspackage}
\usepackage{atlasbiblatex}

\usepackage{atlasphysics}
\graphicspath{{logos/}{figures/}}

\usepackage{ANA-HION-2025-01-PAPER-defs}

\AtlasTitle{Measurement of coherent exclusive $\jpsi\to\mumu$ production in ultraperipheral Pb+Pb collisions at $\sqrt{s_{\textrm{NN}}}=5.36$~TeV with the ATLAS detector}

\AtlasAbstract{%
The ATLAS experiment has performed a  measurement of coherent exclusive $\jpsi\to\mumu$
production in ultraperipheral Pb+Pb collisions at
$\sqrt{s_{\textrm{NN}}}=5.36$~TeV. The data was recorded at the Large Hadron Collider (LHC) during 2023,
and corresponds to an integrated luminosity of 79 $\mu$b$^{-1}$. Exclusive \jpsi candidates
were selected with a dedicated track-sensitive trigger based on the
ATLAS transition radiation tracker.
The
analysis involves reconstruction of the dimuon invariant mass based
on muon tracks from the inner detector, as the muon transverse
momentum range of interest precludes the use of the standard muon
reconstruction and identification algorithms.  Differential
cross sections are measured as a function of $\jpsi$ rapidity and
are compared with theoretical predictions. After extrapolation to
$\sqrt{s_{\textrm{NN}}}=5.02$~TeV, they are also compared with
previous measurements performed by other experiments using data from
LHC Run 2.  While the results agree reasonably well with
theoretical predictions, they are in tension with previous
Run-2 results for the central rapidity region.  }

\AtlasRefCode{HION-2025-01}

\PreprintIdNumber{CERN-EP-2025-180}

\AtlasJournalRef{JHEP 04 (2026) 020}
\AtlasDOI{10.1007/JHEP04(2026)020}

\AtlasCoverEgroupAnalysisTeam{atlas-hion-2025-01-analysis-team@cern.ch}

\hypersetup{pdftitle={ATLAS document},pdfauthor={The ATLAS Collaboration}}

\begin{document}

\maketitle

\tableofcontents

%


\section{Introduction}
\label{sec:Intro}

In ultra-relativistic heavy-ion collisions, the fully-stripped ions
are accompanied by large electromagnetic (EM) fields, which correspond
to large equivalent photon fluxes, often described in the
Weizs\"acker--Williams
approach~\cite{vonWeizsacker:1934nji,Williams:1934ad}.  For photons
emitted coherently by the entire nucleus, the flux is enhanced
compared with beams of protons by a factor of $Z^2$, where $Z$ is the
charge of the nucleus. At large impact parameters, beyond twice the
nuclear radius, hadronic processes are suppressed, and photon-induced
reactions become the dominant interaction mechanism.
Such collisions, referred-to as ultra-peripheral collisions (UPCs),
have been used to study photon--nucleus (photonuclear) and photon--photon
collisions at the Relativistic Heavy Ion Collider (RHIC)~\cite{Harrison:2003sb}
and the Large Hadron Collider (LHC)~\cite{Evans:2008zzb}\@.
Hard-scattering processes in
photonuclear interactions provide a novel method for probing nuclear
parton distributions in a kinematic region not easily accessible to
other measurements~\cite{HION-2022-15}.  In particular,
exclusive diffractive photoproduction of heavy vector mesons is a well-known
process that offers unique access to the spatial and momentum
structure of the nucleon and
nucleus~\cite{Baltz:2002pp,Mantysaari:2022sux}.  UPC events typically
have features such as rapidity gaps, reduced multiplicity, or
exclusive final states that make them qualitatively different from the
typical hadronic nuclear collisions that are used to study ultra-hot
and ultra-dense nuclear matter.

\begin{figure}[b]
\begin{center}
\includegraphics[width=0.5\textwidth]{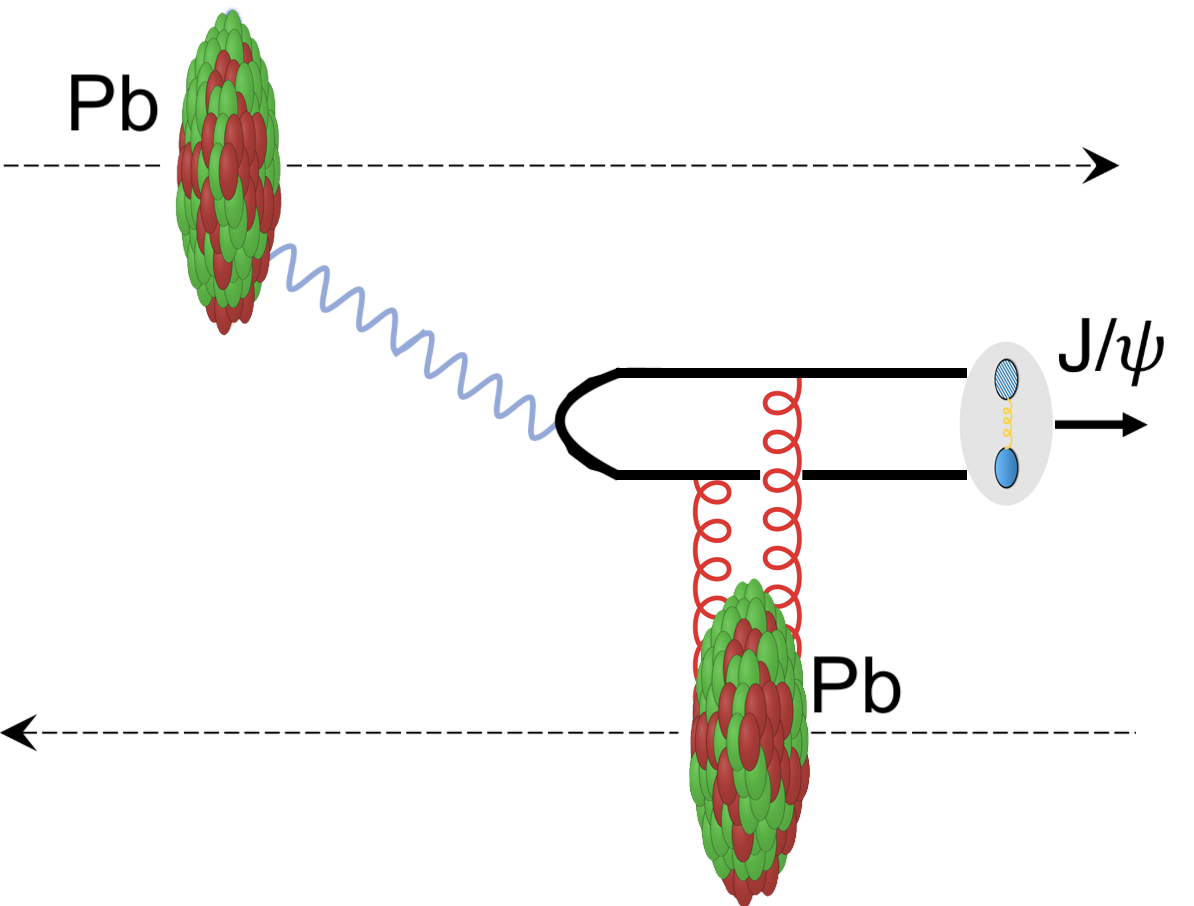}
\caption{Schematic representation of coherent $\jpsi$ meson
photoproduction in Pb+Pb UPC, where a nearly-real photon is emitted
from one nucleus, fluctuates to a charm-anticharm dipole, and is put
on-shell via scattering from a color-neutral pomeron.  }
\label{fig:feyn}
\end{center}
\end{figure}

The physics and phenomenology of UPCs, in particular vector meson
photoproduction, are reviewed in
Refs.~\cite{Baltz:2007kq,Klein:2020fmr}.
Exclusive vector meson photoproduction can proceed in two
ways: coherent or incoherent.
In coherent photoproduction, the
color-neutral exchanged object, typically referred to as a
``pomeron'', is emitted coherently by all the nucleons in the nucleus
and does not directly resolve the nuclear structure, although measurements of this
process provide sensitivity to the average nuclear parton distribution
function (nPDF).
A schematic diagram of the coherent diffractive $\jpsi$
meson photoproduction in UPC is shown in Figure~\ref{fig:feyn}.
In contrast, in incoherent photoproduction, the pomeron is emitted by an
individual nucleon, leading to nuclear breakup, or sometimes resolves
the nucleon's partonic constituents via dissociative production.
Coherent and incoherent exclusive processes can be differentiated experimentally using the magnitude of the
transverse momentum ($\pT$) of the produced vector meson, as this quantity is
inversely related to the transverse size of the ``target'', i.e. the object struck by the photon: coherent
photoproduction involves the entire nucleus, and typically results in vector mesons with low transverse
momentum (of about 50~MeV), while incoherent photoproduction off of nucleons yields
mesons with higher transverse momenta (around 250~MeV for nucleons and
600~MeV for dissociative production).  In the case of coherent
interactions, the nuclei generally remain intact, although the strong
EM fields of ultrarelativistic heavy nuclei can induce additional
breakup through Coulomb dissociation, causing one or both nuclei to
become excited and emit forwardgoing nucleons.  Conversely, incoherent
photoproduction typically leads to nuclear breakup, sometimes with the
emission of additional forward hadrons, which can be misidentified as
an exclusive process due to limited detector acceptance.

Coherent exclusive photoproduction of heavy vector mesons (for example, the
$\jpsi$ meson) is particularly interesting because, at leading order
(LO) in quantum chromodynamics (QCD), the cross section is predicted
to scale approximately with the square of the gluon PDF in the target~\cite{Ryskin:1992ui}.
The scale of the four-momentum transfer of the interaction is related
to the mass ($m_V$) of the vector meson as $Q^2 \approx m_V^2 / 4$,
which implies that the production of heavy charmonium states can be
calculated perturbatively.  Due to the exclusivity of the reaction,
the rapidity of the coherently-produced vector meson is directly
related to the Bjorken-$x$ of the gluons as
$x=m_V/\sqrt{s_{\textrm{NN}}} e^{\pm y}$, where $y$ is the rapidity of
the vector meson and the two possible signs in the exponent reflect
that each of the incoming lead nuclei may act as the photon emitter.
At low Bjorken-$x$ values, the nuclear PDFs are well-known to be
suppressed compared with those of free protons, a phenomenon referred to
as nuclear shadowing~\cite{Armesto:2006ph}.  This suppression is
generally linked to the multiple scattering effects of a projectile in a
nuclear target, and has been explained in various phenomenological
approaches~\cite{Frankfurt:2011cs, Guzey:2013xba, Goncalves:2014wna,Cepila:2017nef,
Bendova:2020hbb, Luszczak:2019vdc}.
Parton saturation phenomena are also expected to emerge at low $x$.
For heavy nuclei such as lead,
due to the recombination of gluons at a sufficiently high density,
they also reduce the rate of $\jpsi$ production (similar to shadowing),
and lead to the
emergence of a new momentum scale, $Q_s$~\cite{PhysRevD.68.114005}.
Heavy vector meson photoproduction measurements thus provide a
powerful tool to study shadowing and saturation phenomena.
However, the sum of the LO and next-to-leading-order (NLO) gluon contributions
exhibits significant cancellations, resulting in a relative dominance of the quark contributions~\cite{Eskola:2022vpi}.
As a result, this process receives sizable NLO corrections, which complicates the interpretation of coherent exclusive $\jpsi$ production data in terms of gluon shadowing.

Charmonium photoproduction in Pb+Pb UPC has been studied widely
by experiments at RHIC~\cite{STAR:2023vvb} and the LHC\@.
Both the ALICE~\cite{ALICE:2012yye, ALICE:2013wjo, ALICE:2015nmy}
and CMS Collaborations~\cite{CMS-HIN-12-009} have performed measurements
at a nucleon--nucleon (NN) center-of mass energy
of $\sqrt{s_{\textrm{NN}}}=2.76$~TeV, in complementary $\jpsi$ rapidity regions.
ALICE has performed measurements near midrapidity ($|y|<0.8$) with both
dimuon and dielectron final states, and at forward rapidities
$2.5 < y < 4$ with dimuons.
CMS has measured the dimuon final state with $1.6<|y|<2.4$.
In LHC Run-2, with $\sqrt{s_{\textrm{NN}}}=5.02$~TeV, updated measurements
were made by ALICE~\cite{ALICE:2019tqa, ALICE:2021gpt, ALICE:2023jgu}
CMS~\cite{CMS-HIN-22-002}, and LHCb~\cite{LHCb:2021bfl,LHCb:2022ahs},
the latter in the forward region $2 < y < 4.5$.

This paper presents the measurement of exclusive
$\jpsi\to\mumu$ photoproduction cross sections in Pb+Pb UPC at
$\sqrt{s_{\textrm{NN}}}=5.36$~TeV, using 79 $\mu$b$^{-1}$ of Run-3 data
recorded in 2023.  The muons are expected to lose all of their energy
in the ATLAS calorimeter material before reaching the muon detectors.
Moreover, electrons from the $\jpsi\to\ee$ process are not considered as signal due to the poor electron energy resolution and identification efficiency in the relevant kinematic range.
As a result, no specific muon or electron identification criteria are applied.
Instead, the analysis relies on the reconstruction of a dimuon invariant
mass
from a pair of opposite-sign charged particles that are assumed to be muons.
After removal of all known backgrounds, the $\jpsi$ cross sections are
measured differentially as a function of rapidity within $|y|<2.5$.
This new measurement complements the previous Run-2 measurements
by covering a previously-unmeasured region $0.8<|y|<1.6$, while also
providing measurements in two regions where previous results exist:
$|y|<0.8$ overlapping with ALICE, and $1.6<|y|<2.4$ overlapping CMS\@.
The measured cross sections are compared with theory predictions,
particularly those which incorporate parton saturation at low Bjorken-$x$,
and the measurements performed at $\sqrt{s_{\textrm{NN}}}=5.02$~TeV by the
other Collaborations,
after a model-dependent extrapolation to account for the beam energy difference.
%


%

%


\section{Experimental configuration}
\label{sec:atlas}

\newcommand{\AtlasCoordFootnote}{%
ATLAS uses a right-handed
coordinate system with its origin at the nominal interaction point
in the centre of the detector and the \(z\)-axis along the beam
pipe.  The \(x\)-axis points from the IP to the centre of the LHC
ring, and the \(y\)-axis points upwards.  Polar coordinates
\((r,\phi)\) are used in the transverse plane, \(\phi\) being the
azimuthal angle around the \(z\)-axis.  The pseudorapidity is
defined in terms of the polar angle \(\theta\) as \(\eta = -\ln
\tan(\theta/2)\) and is equal to the rapidity $ y = \frac{1}{2} \ln
\left( \frac{E + p_z}{E - p_z} \right) $ in the relativistic limit.
Angular distance is measured in units of \(\Delta R \equiv
\sqrt{(\Delta y)^{2} + (\Delta\phi)^{2}}\).
}

The ATLAS experiment~\cite{PERF-2007-01,GENR-2019-02} at the LHC is a
multipurpose particle detector with a forward--backward symmetric
cylindrical geometry and a near \(4\pi\) coverage in solid
angle.\footnote{\AtlasCoordFootnote} It consists of an inner tracking
detector (ID) surrounded by a thin superconducting solenoid providing
a \qty{2}{\tesla} axial magnetic field, electromagnetic and hadronic
calorimeters, and a muon spectrometer.  The ID covers the
pseudorapidity range \(|\eta| < 2.5\).  It consists of silicon pixel,
silicon microstrip, and transition radiation tracking (TRT) detectors.
The TRT is the outermost part of the ATLAS ID, and it measures charged
particle trajectories before they impinge upon the ATLAS calorimeters.
It is divided into four $\eta$ regions:
barrel A ($0< \eta \lesssim 0.8$),
barrel C ($-0.8 \lesssim \eta<0$),
endcap A $(0.8 \lesssim \eta\lesssim 2.0$)
and endcap C $(-2.0\lesssim \eta \lesssim -0.8$).
The inner radius of the TRT of about 600 mm makes it sensitive to low-\pT charged particles,
nominally down to $\pt \approx 0.5$~GeV, below which the curvature of tracks in the ATLAS
solenoidal field prevents them from reaching it.
Lead/liquid-argon (LAr) sampling calorimeters provide EM energy
measurements with high granularity within the region \(|\eta|< 3.2\).
A steel/scintillator-tile hadronic calorimeter covers the central
pseudorapidity range (\(|\eta| < 1.7\)).  The endcap and forward
(FCal) regions are instrumented with LAr calorimeters for EM and
hadronic energy measurements up to \(|\eta| = 4.9\).  The muon
spectrometer surrounds the calorimeters and is based on three large
superconducting air-core toroidal magnets with eight coils each.  The
field integral of the toroids ranges between \num{2.0} and
\qty{6.0}{\tesla\metre} across most of the detector. The muon
spectrometer includes a system of precision tracking chambers up to
\(|\eta| = 2.7\) and fast detectors for triggering up to \(|\eta| =
2.4\).  The luminosity is measured mainly by the LUCID--2 detector
which is located close to the beampipe.  Forward neutrons produced from
the breakup of lead nuclei in both hadronic and electromagnetic interactions
are measured by compact tungsten sampling Zero Degree Calorimeters (ZDCs) positioned
at $z=\pm140$ meters from the ATLAS interaction point.

A two-level trigger system is
used to select events~\cite{TRIG-2022-01}.  The first-level (L1) trigger  is
implemented in hardware and uses a subset of the detector information
to accept events at a rate close to \qty{100}{\kHz}.  This is followed
by a software-based trigger that reduces the accepted rate of complete
events to \qty{3}{\kHz} on average, depending on the data-taking
conditions.  A software suite~\cite{SOFT-2022-02} is used in data
simulation, in the reconstruction and analysis of real and simulated
data, in detector operations, and in the trigger and data acquisition
systems of the experiment.

As the final state of $\jpsi\to\mumu$ consists of two soft muons, each
with a maximum \pT just over 1.5~GeV, the analysis primarily makes use
of the ID.
In particular, to trigger on the desired final state with two low momentum tracks, the TRT
``FastOR'' trigger is used at L1~\cite{Fratina:1229213}.

\begin{figure}[t!]
\begin{center}
\includegraphics[width=0.7\textwidth]{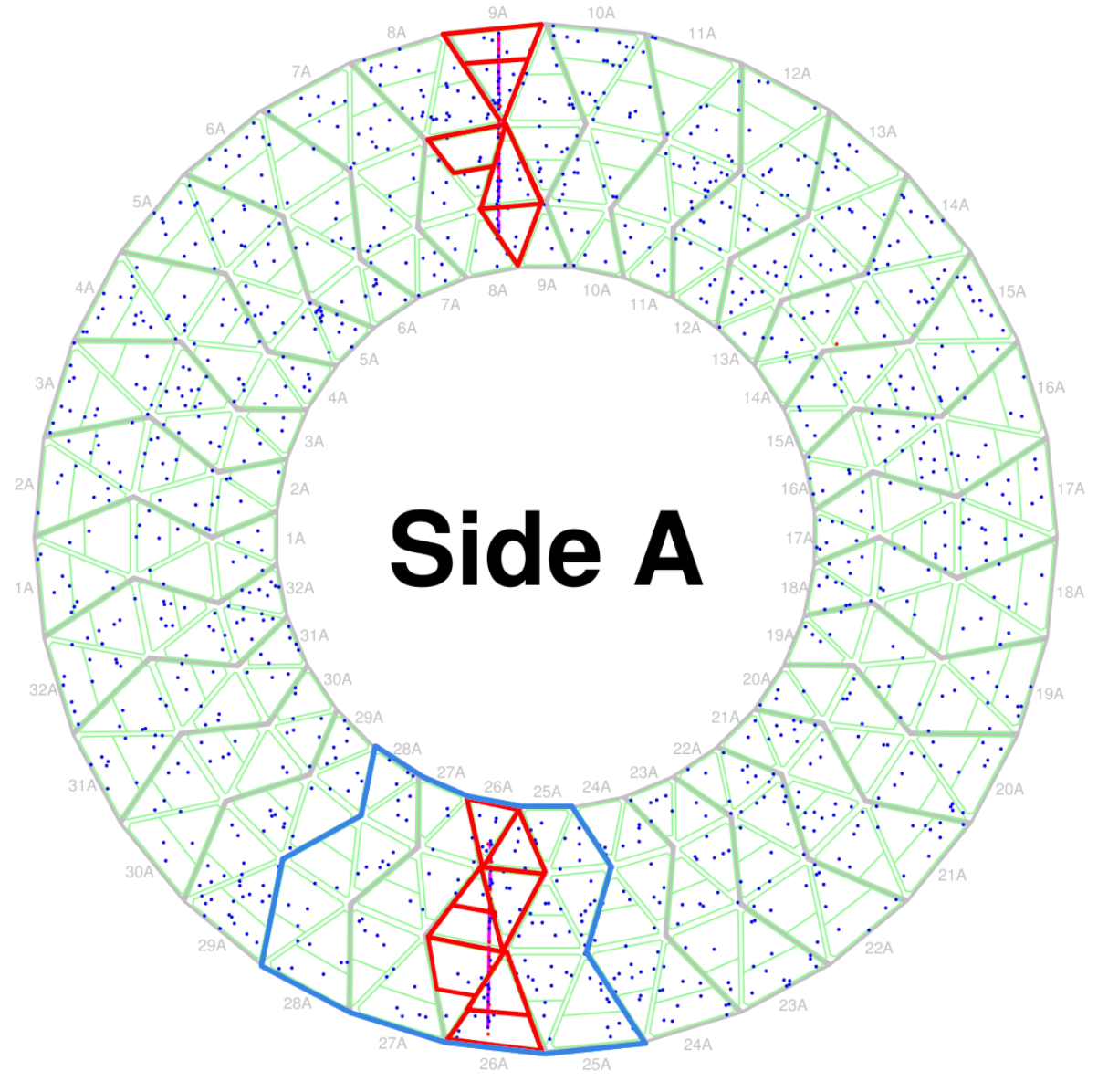}
\caption{ The L1 TRT trigger logic in the TRT barrel A
region ($0< \eta \lesssim 0.8$)~\cite{Fratina:1229213}.  The region is divided
into 32 $\phi$ sectors (bold gray), each consisting of nine TRT trigger
segments (shown as green triangles and trapezoids). A trigger signal
within a segment is formed if the transition radiation threshold is
exceeded in any of TRT straws from this segment (shown as red
triangles and trapezoids, induced by a vertical cosmic track in this event).
The trigger logic then aggregates signals
from all segments in a region formed from four adjacent $\phi$ sectors
(shown in blue), yielding an effective $\phi$
segmentation of $2\pi/8$. To form a trigger decision for this TRT section,
the number of segments with a trigger signal has to exceed a fixed threshold
multiplicity ($M=4$) in at least one contiguous set of four sectors.  }
\label{fig:fastor}
\end{center}
\end{figure}

The L1 TRT trigger logic is illustrated in Figure~\ref{fig:fastor} for the TRT barrel A region.
The trigger is derived from a logical OR of fast electronic
signals, reflecting the passage of tracks through the straw tubes.
The TRT straw tubes are connected to the readout electronics
via front-end modules which each process signals from 16 straws.
Each TRT front-end module can be configured to send a signal that is
generated using a fast digital OR circuit (FastOR) to the TRT trigger
timing and control (TTC) board if a hit exceeding a high threshold level
(HT) is recorded in at least one of its associated straws. A logic
circuit on the TTC board aggregates these signals from groups of
several front-end chips, which reflect contiguous azimuthal regions.
If the number of received signals within a
TTC board exceeds a configurable threshold, set to four in 2023 Pb+Pb collisions,
a trigger signal is generated for the region.
All four TRT regions provide a similar binary trigger
signal, which are logically OR'ed to provide a fast binary trigger decision for each event.

Until 2023, the L1 TRT trigger had been used primarily for cosmic ray
triggering in ATLAS\@.  For the 2023 heavy-ion run, it was re-deployed
with some operational modifications to provide triggering on final
states with up to $\mathcal{O}(10)$ low-$\pT$ tracks.  Due to high hit
occupancy, the L1 TRT trigger is unsuitable for nominal proton--proton
($pp$) collisions and is only viable in low-pileup conditions such
as cosmic-ray or Pb+Pb runs.  Typically, the TRT HT level is
calibrated to the transition radiation threshold and is used for
electron identification~\cite{IDET-2015-01}.
In Pb+Pb collisions, this was lowered to about twice the low threshold
used in the standard $pp$ tracking algorithms~\cite{IDET-2015-01}, such
that a typical charged particle reaching the TRT can produce a large
number of HT hits. In this way, the TRT was able to
trigger on events with even just one or two low-momentum tracks.

%

%


\section{Data and Monte Carlo simulation samples}
\label{sec:Datamc}

The data used in this measurement were recorded by ATLAS in 2023 during Pb+Pb operation of the LHC at $\sqrt{s_{\textrm{NN}}}=5.36$~TeV.
To accumulate $\jpsi\to\mumu$ events with a good efficiency while rejecting hadronic
heavy ion collisions, candidate events were recorded using a dedicated trigger for
events with moderate activity in the ID but little additional activity
in the entire detector.  The L1 decision required both a signal
from the L1 TRT FastOR system  and a total transverse energy
(\et) below 20~GeV (at electromagnetic scale) registered in the
calorimeter system.  At the HLT, the total \et in the FCal region on
each side was required to be below 5~GeV, to suppress hadronic events.
In addition, the ID was used at the HLT to select events with up
to 15 tracks with $\pT > 100$~MeV, with an additional requirement of
1--5 tracks with $\pT>1$~GeV.  The efficiency of the trigger
selection is described in Section~\ref{sec:trigger_eff}.
The trigger was prescaled at L1 during the data-taking period (meaning that not every event that satisfied the trigger requirements was recorded for further processing).
The data set corresponds to an integrated luminosity of 79 $\mu$b$^{-1}$.

Monte Carlo (MC) simulated events for the coherent UPC $\jpsi\to\mumu$
process were produced using version 2.0 of the STARlight
generator~\cite{Klein:2016yzr} interfaced with
\PYTHIA~8.312~\cite{Bierlich:2022pfr} to model final-state radiation
(FSR) from the muons~\cite{HION-2016-02}.  For the $\jpsi\to\mumu$
process, STARlight relies on the vector meson dominance (VMD) model
and utilizes a parameterization of existing data on $\jpsi$
photoproduction off protons.  To determine the $\jpsi$ photoproduction
cross section in Pb+Pb UPC, a Glauber formalism is applied that
considers multiple interactions within the nucleus but does not
include explicit gluon-shadowing corrections.

The sources of background in this analysis include coherent
$\Jpsi\to\ee$, incoherent $\jpsi\to\ellell$, coherent and incoherent
$\psip$, and $\gamma\gamma\to \ellell$ processes, where $\ell=e,\mu$.
These backgrounds were also modeled with STARlight.  It was similarly
interfaced to \PYTHIA~8 for coherent $\jpsi\to\ee$, incoherent
$\jpsi\to \ellell$ and $\gamma\gamma\to\ellell$ UPC production.  The
feed-down background from the production of coherent and incoherent
$\psip$ vector meson in UPC was generated with STARlight, and
\EVTGEN~2.2.1~\cite{Lange:2001uf} was used to provide all known
$\psi(2S)$ decay channels.  The FSR from the charged decay products of
$\psip$ vector mesons was simulated with
\PHOTOSpp~3.61~\cite{Davidson:2010ew}.

In all of the STARlight events, the produced $\jpsi$ and
$\psip$ vector mesons were assumed to be transversely
polarized. This is consistent with expectations from helicity
conservation in photoproduction processes and is also consistent with experimental
observations~\cite{ALICE:2023svb}.
The generated \pT distribution of coherent $\jpsi$ from STARlight is reweighted to match the measured cross sections by the LHCb Collaboration~\cite{LHCb:2022ahs}.
The measured coherent $\jpsi$ \pT shape from LHCb is found to be consistent with the shape measured by ALICE at mid-rapidity~\cite{ALICE:2021tyx}.
This reweighting procedure leads to corrections at the level of approximately 10–20\% in the $\jpsi$ \pT spectrum and significantly improves the modeling of the $\jpsi$ \pT shape in the entire rapidity range.

%
%

%
%
%
%
%
%
%
%
%
%
%
%
%
%
%
%
%
%
%
%
%
%
%
%
%
%
%
%
%
%
%
%


%

%


\section{Signal selection}
\label{sec:Evtsel}

The selected signal events are required to have exactly two,
opposite-charge tracks.  Charged-particle tracks are reconstructed in
the ID using an algorithm optimized for $pp$ minimum-bias
measurements~\cite{ATL-PHYS-PUB-2015-018}.  They are pre-selected to
have $\pT>100$~\MeV and $|\eta|<2.5$, and to have a transverse impact
parameter calculated relative to the measured beam-line position of
$|d_0|<2$~mm.  The Loose Primary track
selection~\cite{ATL-PHYS-PUB-2015-051,PERF-2015-07,PERF-2015-08} is
used to reject poorly-reconstructed tracks.

Apart from the track veto (exactly two tracks with $\pT>100$~MeV and $|\eta|<2.5$), no other exclusivity selection criteria are applied. However, various cross-checks are performed (see Section~\ref{sec:sysother}) to ensure that the selected events are produced without additional hadronic activity.
These cross-checks also ensure that the applied exclusivity selection criteria are insensitive to potential biases arising from additional UPC interactions occurring in the same or in separate Pb+Pb collisions.

In addition, the \pT of
both tracks is required to be above 1~GeV to coincide with the trigger
requirements.  The muon mass hypothesis is used for each track.
Due to the very low \pT of the muons from the $\jpsi$ decay, no specific identification criteria are used, as these particles are expected to lose all their energy and stop in the calorimeter before they can reach the muon detectors.
The signal region (SR) selection then requires the invariant mass of
the two-track system to be in the range of $2.9<m_{\trktrk}<3.2$~GeV,
and the pair transverse momentum to be $\pT^{\trktrk}<0.2$~GeV.
A small $\pT^{\trktrk}$ is required because coherent photoproduction involves the entire nucleus and thus imparts very little transverse momentum to the \jpsi.
%


%

%


\section{Background estimate and signal yield extraction}
\label{sec:yield}

This section summarizes the estimations of the primary background
sources, and the signal yield extraction procedure.  The main sources
of background are the $\gamma\gamma\to \ellell$ and $\Jpsi\to\ee$
processes.  Other sources of background are the incoherent
$\jpsi\to\ellell$ process, coherent and incoherent $\psip$ production,
and a combinatorial background of pion pairs from inelastic
photonuclear processes.

\subsection{$\psip$ feed-down}
\label{sec:psi2s}
Photoproduced $\psip$ mesons can decay to $\jpsi+X$, where $X$ denotes
additional particles (typically soft pions), which can go undetected
due to the limited ID acceptance in $\eta$ or $\pT$, and
material interactions that preclude their reconstruction.  Yields of
these events are estimated with MC simulations, with an additional
normalization factor extracted from a control region.  This control region selects events with a $\psip \to
\jpsi(\to\ellell)\pi^+\pi^-$ decay and it requires:
\begin{itemize}
\item exactly three or four tracks, where the leading
(subleading) track has $\pT>1$~GeV ($\pT~>~0.8$~GeV) and
the two leading tracks have opposite charge;
\item the third-leading track has $\pT>0.2$~GeV;
\item if exactly three tracks are present, exactly one pixel
track\footnote{Pixel tracks are reconstructed using hits from the
pixel detector only. They are required to have $\pT > 50$~MeV,
$|\eta|<2.5$, and at least three hits in the pixel detector. As
pixel tracks provide much better reconstruction efficiency at
$\pT\sim 100$~MeV relative to full ID tracks, and because the
pions from $\psip \to \jpsi(\to\ellell) \pi^+\pi^-$ are very soft,
this procedure significantly improves the $\psip$ selection efficiency.}
unmatched with the other three tracks (with a
$\Delta R(\textrm{track,~pix. track})<0.4$ matching condition) is required, and is also called a \textit{track} in the following;
then, the 3rd and 4th leading tracks are also required to have
opposite charge;
\item the invariant mass of the two
leading tracks to satisfy  $3<m_{\trktrk}<3.2$~GeV, and
the invariant mass of the full system to satisfy
$3.5<m_{\trktrk\pi^+\pi^-}<3.8$~GeV, where the charged pion mass hypothesis is used for the 3rd and 4th leading tracks;
\item the acoplanarity ($A_{\phi}=1-|\Delta\phi|/\pi$) between the leading and subleading track system, and the 3rd and 4th leading track system, $A_{\phi}(\trktrk,\pi^+\pi^-)<0.5$.

\end{itemize}

The dominant background for this $\psip$ decay process
comes from combinatorial processes (multiple
pion production) and is estimated by using events from $m_{\trktrk}$
low-mass ($2.6<m_{\trktrk}<2.75$~GeV) and high-mass
($3.25<m_{\trktrk}<3.4$~GeV) sidebands.
Other background includes $\jpsi\rho^0$
production, and it is estimated by using a dedicated control region. First, the
$A_{\phi}(\trktrk,\pi^+\pi^-)<0.5$ requirement is inverted. This is because the $\jpsi\rho^0$ background is flat with
$A_{\phi}(\trktrk,\pi^+\pi^-)$, whereas $\psip \to
\jpsi(\to\ellell)\pi^+\pi^-$ events concentrate at low
$A_{\phi}(\trktrk,\pi^+\pi^-)$ values.
Then, additional requirements of $A_{\phi}(\mu^+,\mu^-)<0.02$ and
$A_{\phi}(\pi^+,\pi^-)<0.1$ are used to reflect the fact
that $\jpsi$ and $\rho^0$ decays are from two separate photonuclear processes,
even when produced by the same Pb+Pb interaction.
Moreover, the $3.5<m_{\trktrk\pi^+\pi^-}<3.8$~GeV requirement is relaxed to $m_{\trktrk\pi^+\pi^-}<4$~GeV.

To extract the coherent and incoherent $\psip$ components, a fit to
the four-track system \pT distribution is performed, with the
normalization factors for the two components as free parameters.
These factors are then used to normalize the coherent and incoherent $\psip$ MC samples, which are subsequently used to estimate the $\psip$ background in the SR.

\subsection{Pions from inclusive photonuclear events}
\label{sec:pions}
\begin{figure}[t!]
\begin{center}
\subcaptionbox{\label{fig:pions_cr_a}}{\includegraphics[width=0.495\textwidth]{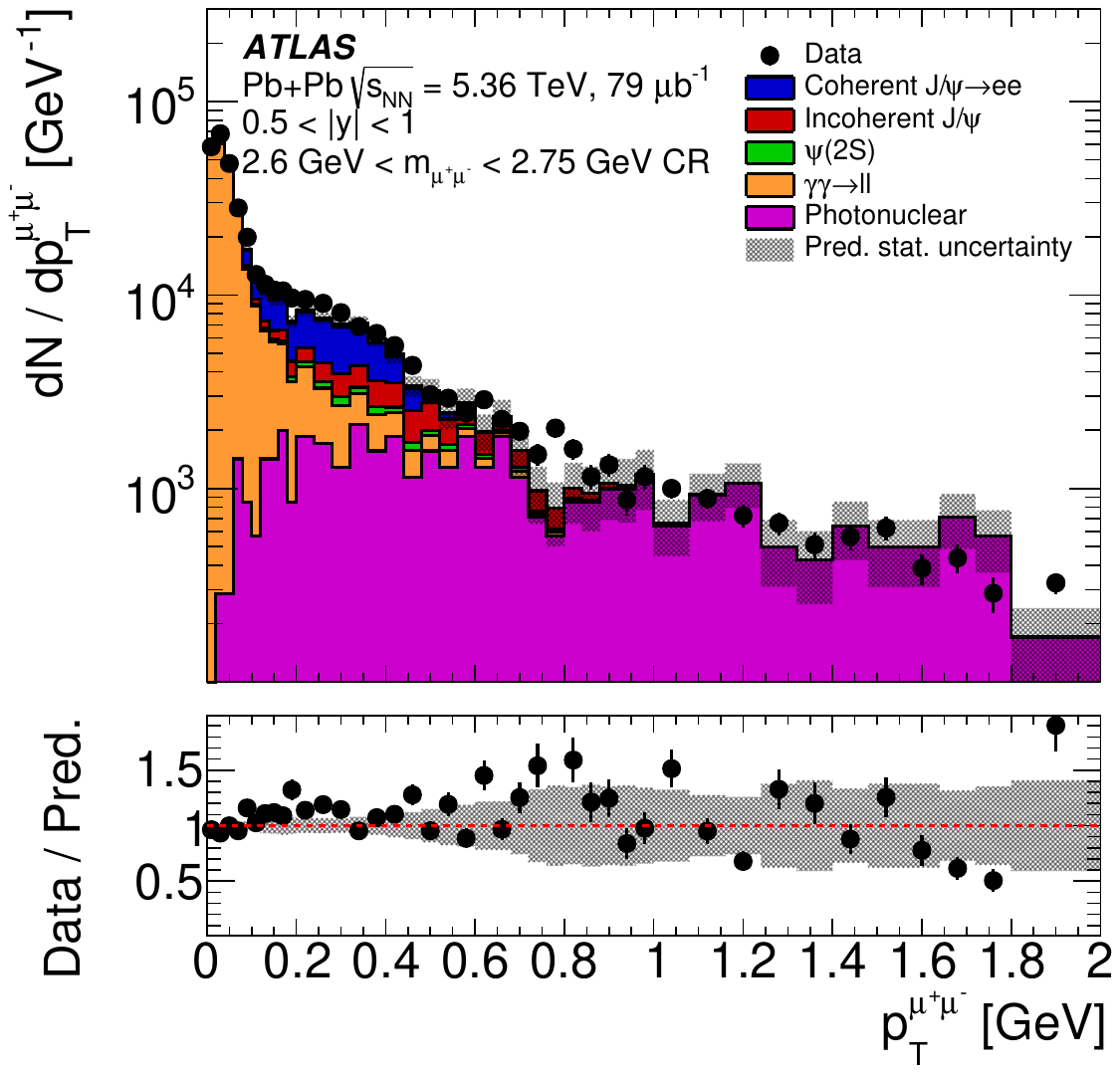}}
\subcaptionbox{\label{fig:pions_cr_b}}{\includegraphics[width=0.495\textwidth]{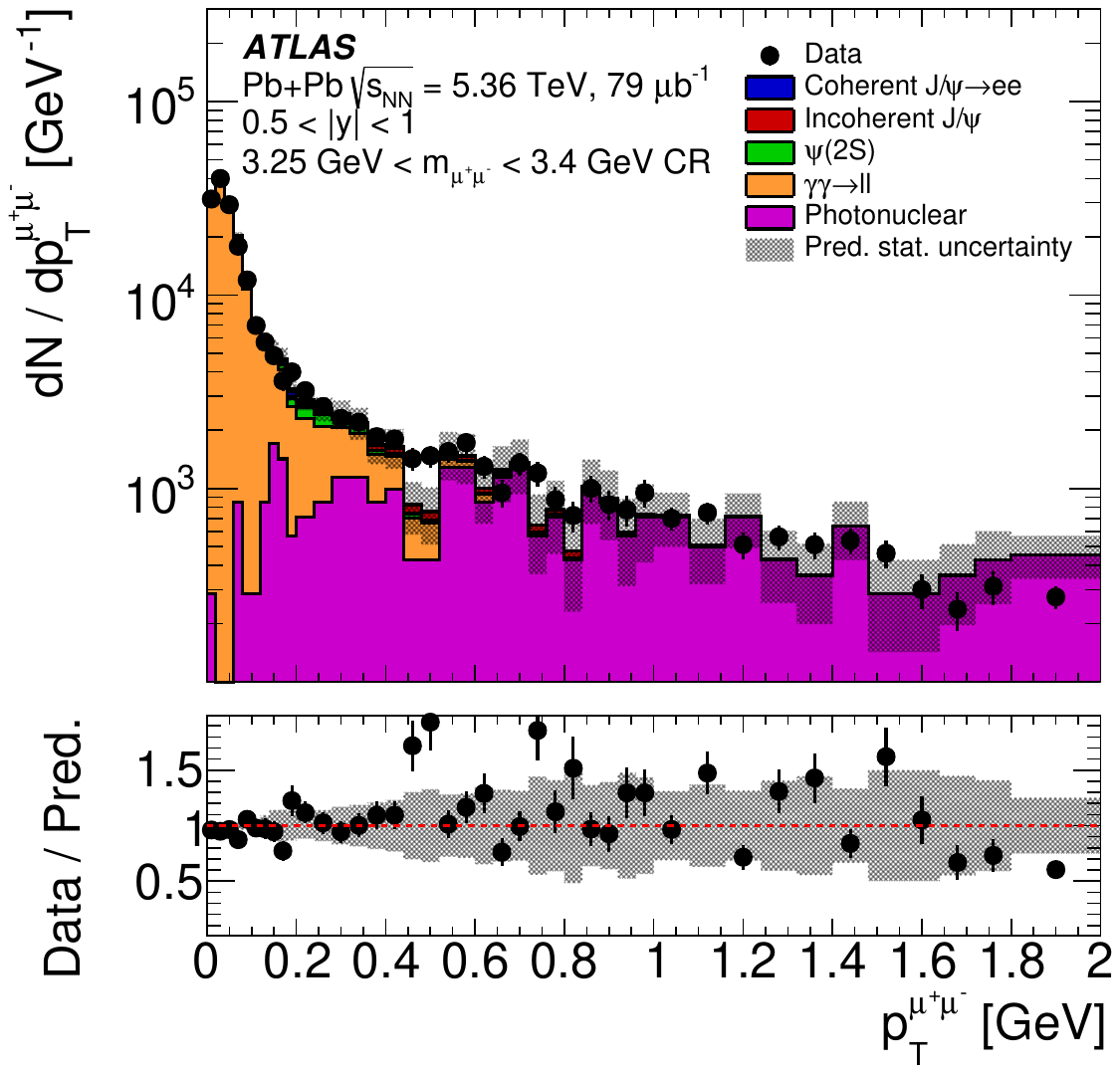}}
\caption{ Observed exclusive two-track system \pT distributions
together with fit signal and background contributions in the
rapidity range $0.5<|y|<1$ for the (a) low-mass
($2.6<m_{\trktrk}<2.75$~GeV) and (b) high-mass
($3.25<m_{\trktrk}<3.4$~GeV) sideband regions.  The photonuclear
background templates are normalized in the low-mass sideband region
in the $\pT^{\trktrk}>1$~GeV range.
The signal contribution (coherent $\jpsi\to\mumu$) is negligible in these two regions and hence it is not shown.
For the calculation of $\pT^{\trktrk}$, the muon mass hypothesis is used for each track.
}
\label{fig:pions_cr}
\end{center}
\end{figure}

As the analysis does not use any specific identification criteria for
muons, a combinatorial background from hadronic
photonuclear events is expected.  In such events, a photon fluctuates into a
hadronic state, typically a $\rho^0$, and interacts with the nucleus to
form a multi-hadron final state.  This can lead to events in which
only a few particles are observed in the ID.
The contribution of these events to the SR is estimated by using a
data-driven procedure using same-sign track pairs.  Template
$\pT^{\mu^{\pm}\mu^{\pm}}$ distributions are obtained from events
satisfying all signal selection criteria except the opposite-sign
requirement, which is inverted.
Templates are derived in the range of $\pT^{\mu^{\pm}\mu^{\pm}}<2$~GeV,
for the SR ($2.9<m_{\trktrk}<3.2$~GeV), and for both the low-mass
($2.6<m_{\trktrk}<2.75$~GeV) and high-mass
($3.25<m_{\trktrk}<3.4$~GeV) sideband regions.  These
photonuclear background templates are then normalized in the low-mass
sideband region for events having $\pT^{\trktrk}>1$~GeV where
other contributions ($\gamma\gamma\to\ellell$, $\jpsi\to\ellell$)
are found to be suppressed.  This normalization is shown in
Figure~\ref{fig:pions_cr_a} for a representative $\jpsi$ rapidity bin.
As a cross-check of the full procedure, the high-mass sideband region is shown in Figure~\ref{fig:pions_cr_b}, where a good
agreement between the data and background estimate is observed, using the results from the
low-mass fit.

\subsection{$m_{\trktrk}$ fits for $\gamma\gamma\to \ellell$ and $\jpsi$ processes}
\label{sec:mass_fit}

\begin{figure}[!htbp]
\begin{center}
\vspace*{-0.5cm}
\subcaptionbox{\label{fig:data_fit_jpsi1}}{\includegraphics[width=0.46\textwidth]{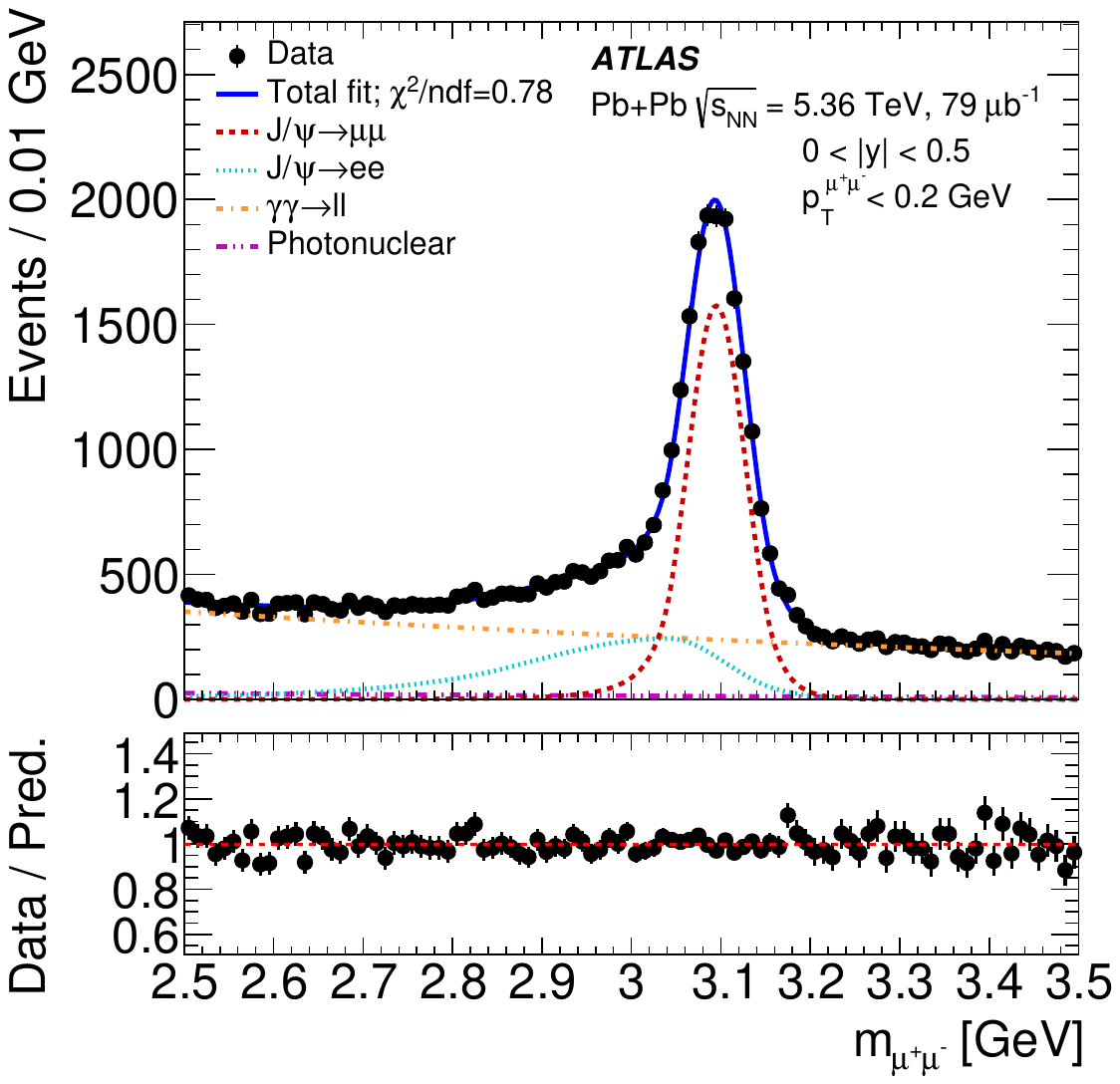}\vspace*{-0.5cm}}
\vspace*{-0.1cm}
\subcaptionbox{\label{fig:data_fit_jpsi2}}{\includegraphics[width=0.46\textwidth]{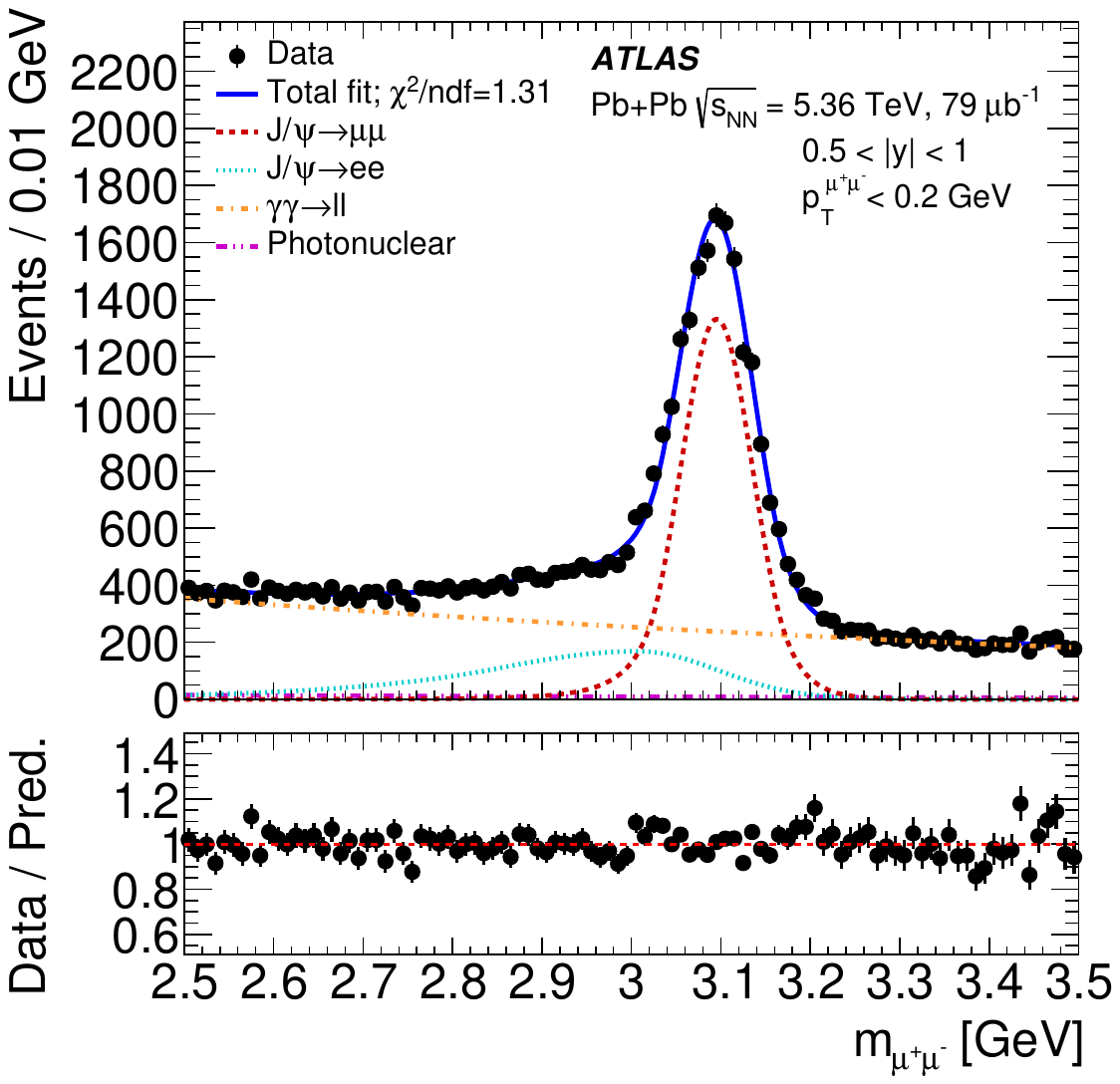}\vspace*{-0.5cm}}
\vspace*{-0.1cm}
\subcaptionbox{\label{fig:data_fit_jpsi3}}{\includegraphics[width=0.46\textwidth]{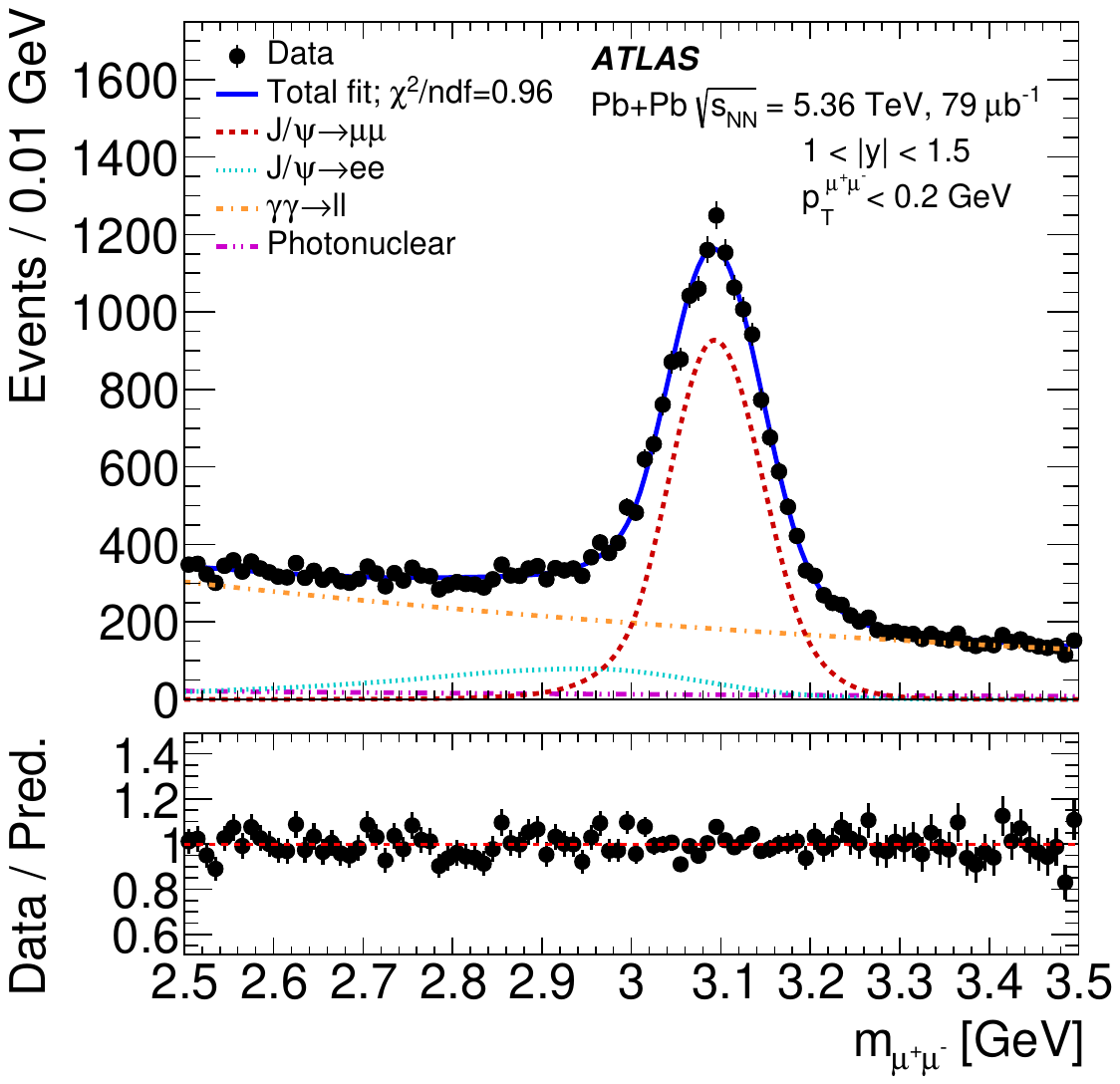}\vspace*{-0.5cm}}
\vspace*{-0.1cm}
\subcaptionbox{\label{fig:data_fit_jpsi4}}{\includegraphics[width=0.46\textwidth]{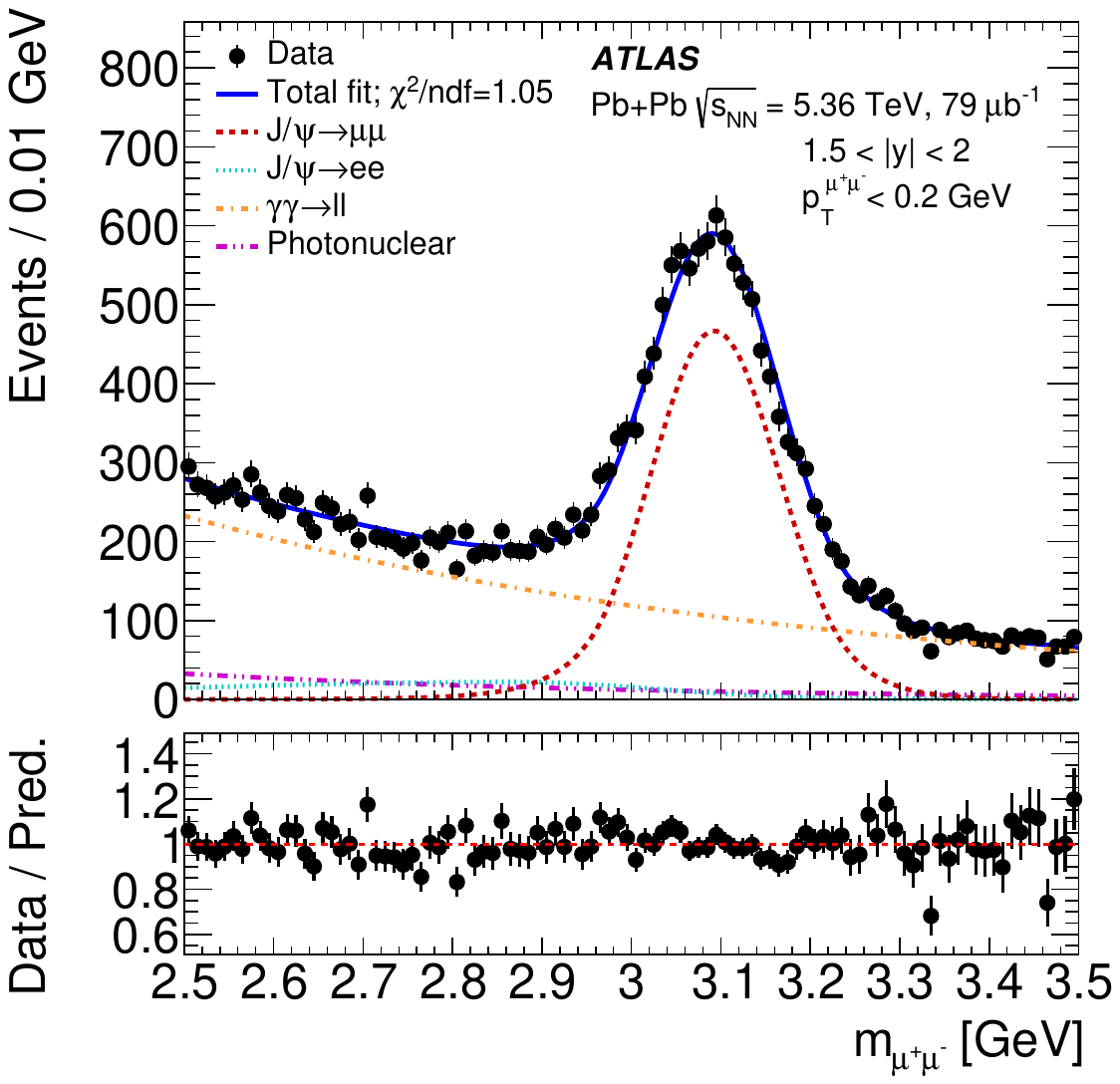}\vspace*{-0.5cm}}
\vspace*{-0.1cm}
\subcaptionbox{\label{fig:data_fit_jpsi5}}{\includegraphics[width=0.44\textwidth]{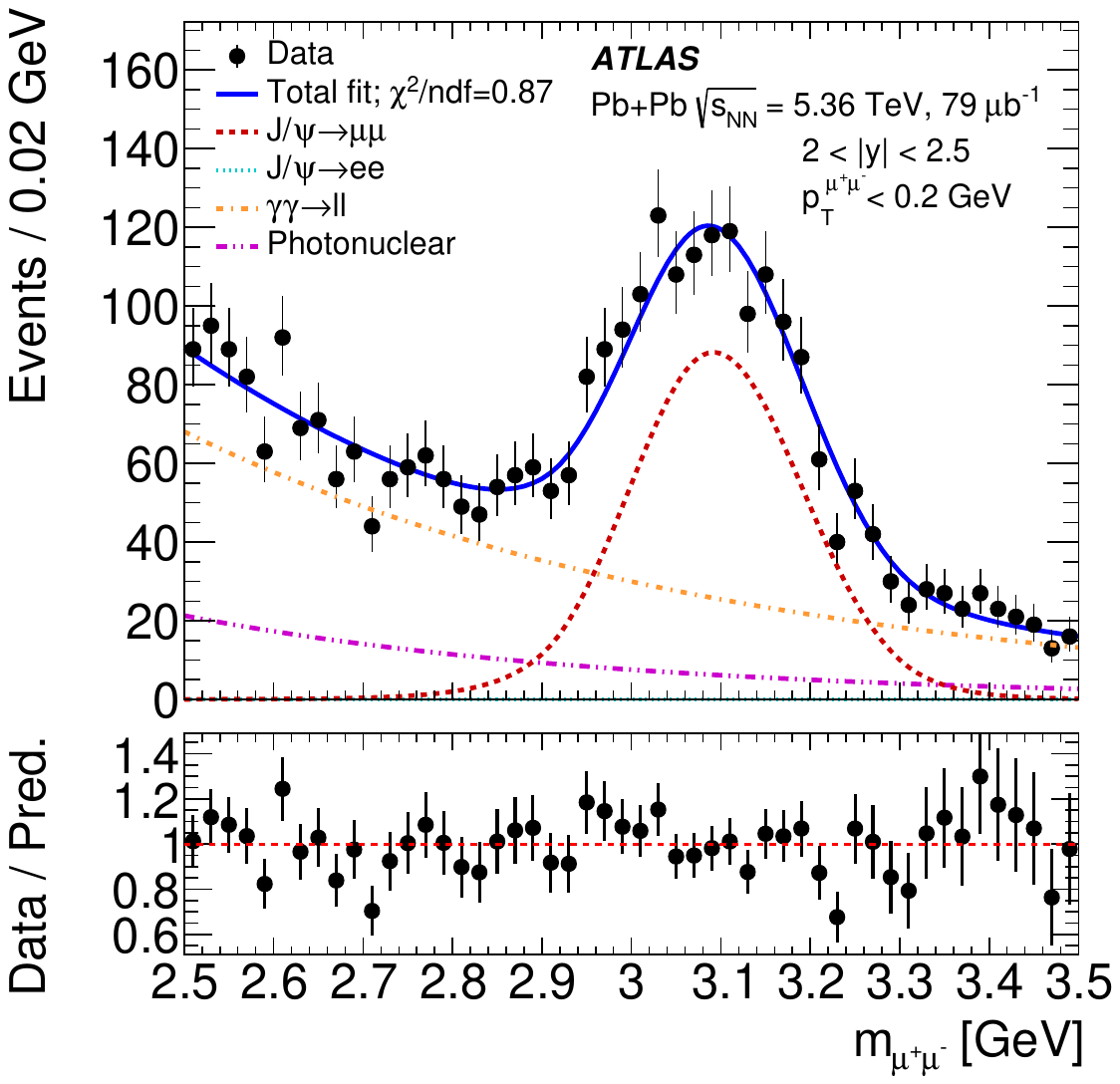}\vspace*{-0.5cm}
}
\caption{ Exclusive two-track invariant mass ($m_{\trktrk}$)
distributions
for five $|y|$ intervals.  Resonant contributions from
$\jpsi\to\mumu$ and $\jpsi\to\ee$ processes are shown as long-dashed and
short-dashed lines, respectively.  Non-resonant contributions
from $\gamma\gamma\to\ellell$ and photonuclear backgrounds are shown
as dotted-dashed and two-dotted dashed lines respectively. The total fit
result for each rapidity region is shown as a solid line. The bottom panel in each figure shows the ratio of data to the total prediction. For
the calculation of $m_{\trktrk}$, the muon mass hypothesis is used
for each track. The signal region is defined as $2.9<m_{\trktrk}<3.2$~GeV.}
\label{fig:data_fit_jpsi}
\end{center}
\end{figure}

Electrons from the $\jpsi \to e^+e^-$ process experience significant detector distortions due to energy loss and radiative processes in the ID material, resulting in broadened and shifted mass peaks compared with the $\jpsi \to \mu^+\mu^-$ decay.
To discriminate between the $\jpsi\to\mumu$, $\jpsi\to\ee$ and
$\gamma\gamma\to\ellell$ processes, template fits to the two-track invariant
mass spectra for SR events in the extended two-track invariant mass
range of $2.5<m_{\trktrk}<3.5$~GeV, are performed.
The $\jpsi\to\ellell$ process considered here includes
a sum of coherent, incoherent and $\psip$ feed-down contributions.
The fits use the following inputs:
\begin{itemize}
\item $\jpsi\to\mumu$ and
$\jpsi\to\ee$ mass shapes described by double-sided Crystal Ball
functions~\cite{Oreglia:1980cs}, whose parameters are
fixed using fits to simulated events,
\item a small contribution from pions from inclusive photonuclear events
modeled using same-sign templates from data, as explained in
Section~\ref{sec:pions}, and further smoothed using an exponential
function ($e^{p_0+p_1\cdot m}$, where $p_0$ and $p_1$ are the fit parameters),
\item an exponential function describing the $\gamma\gamma\to\ellell$ background process, with a
floating slope parameter.
\end{itemize}
In total, this results in four fit
parameters for each rapidity interval: normalization factors for
$\jpsi\to\mumu$ and $\jpsi\to\ee$ contributions and two
parameters for the $\gamma\gamma\to\ellell$ background (slope and normalization).
Binned maximum-likelihood fits are performed in each
two-track rapidity interval, with results shown in
Figure~\ref{fig:data_fit_jpsi}. %
The fits well describe the two-track invariant mass spectra observed in data, with a chi-squared per degree of freedom for each distribution in the range of 0.8--1.3.

The suppression of the $\jpsi \to \ee$ contribution at higher $|y|$ regions, as shown in Figure~\ref{fig:data_fit_jpsi}, is attributed to the increased amount of ID material with $|\eta|$, which affects both the track reconstruction efficiency and the momentum reconstruction for electrons.
It is worth noting that the ratio of the fitted $\jpsi \to \ee$ and $\jpsi \to \mumu$ yields is consistent with the ratio predicted by the MC simulation within 20\% across the entire rapidity range.

To cross-check the absolute rate of the $\gamma\gamma\to\ellell$ background process, events in the low-mass
and high-mass sideband regions are examined.
The $\gamma\gamma\to\ellell$ predictions, normalized using cross sections from STARlight, are below the observed yields by approximately 20\%.
This discrepancy is consistent with previous observations of the $\gamma\gamma\to\ellell$ process in Pb+Pb UPC at low dilepton invariant masses~\cite{ALICE:2013wjo, CMS-HIN-21-015, HION-2021-16}.

\subsection{Dissociative incoherent $\jpsi\to\ellell$ production}

Incoherent $\jpsi\to\ellell$ production with nucleon dissociation is
also taken into account to describe the high-$\pT^{\trktrk}$ tail with
the template based on the parameterization used by the H1 Collaboration~\cite{H1:2013okq}.
The differential proton-dissociative cross section is parameterised, as a function of
$\pT^{\trktrk}$, with a power-law function $d\sigma / d\pT \approx \pT (1 + (b_{pd}/n)\pT^2 )^{-n}$.
The parameters are set to
$n=3.58$ and $b_{pd}=1.79$~GeV$^{-2}$, according to fits to $\gamma
p\to J/\psi X$ data~\cite{H1:2013okq}.  This functional form has also been
used in previous ALICE and CMS
measurements~\cite{ALICE:2019tqa,ALICE:2021gpt, ALICE:2023jgu,CMS-HIN-22-002}.

\subsection{Yield extraction fits to $\pT^{\trktrk}$}

The raw exclusive UPC $\Jpsi$ yields obtained from the invariant mass
fits contain contributions from coherent and incoherent $\Jpsi$
photoproduction that can be separated via the analysis of the
$\pT^{\trktrk}$ spectra.

Individual $\pT^{\trktrk}$ distributions are fit in the signal region
$2.9<m_{\trktrk}<3.2$~GeV with MC templates produced
using the simulated STARlight samples, corresponding to the
production mechanisms discussed above:
coherent and incoherent $\jpsi\to\ellell$,
feed-down $\jpsi\to\ellell$ from decays of coherent and incoherent
$\psip$ and the dilepton continuum from the
$\gamma\gamma\to\ellell$ process (both to dimuons and dielectrons).
In addition, pions from inclusive
photonuclear events are included using the $\pT^{\trktrk}$
templates obtained from same-sign events.

\begin{figure}[!htbp]
\begin{center}
\vspace*{-0.5cm}
\subcaptionbox{\label{fig:incoh_frac_fill1}}{\includegraphics[width=0.46\textwidth]{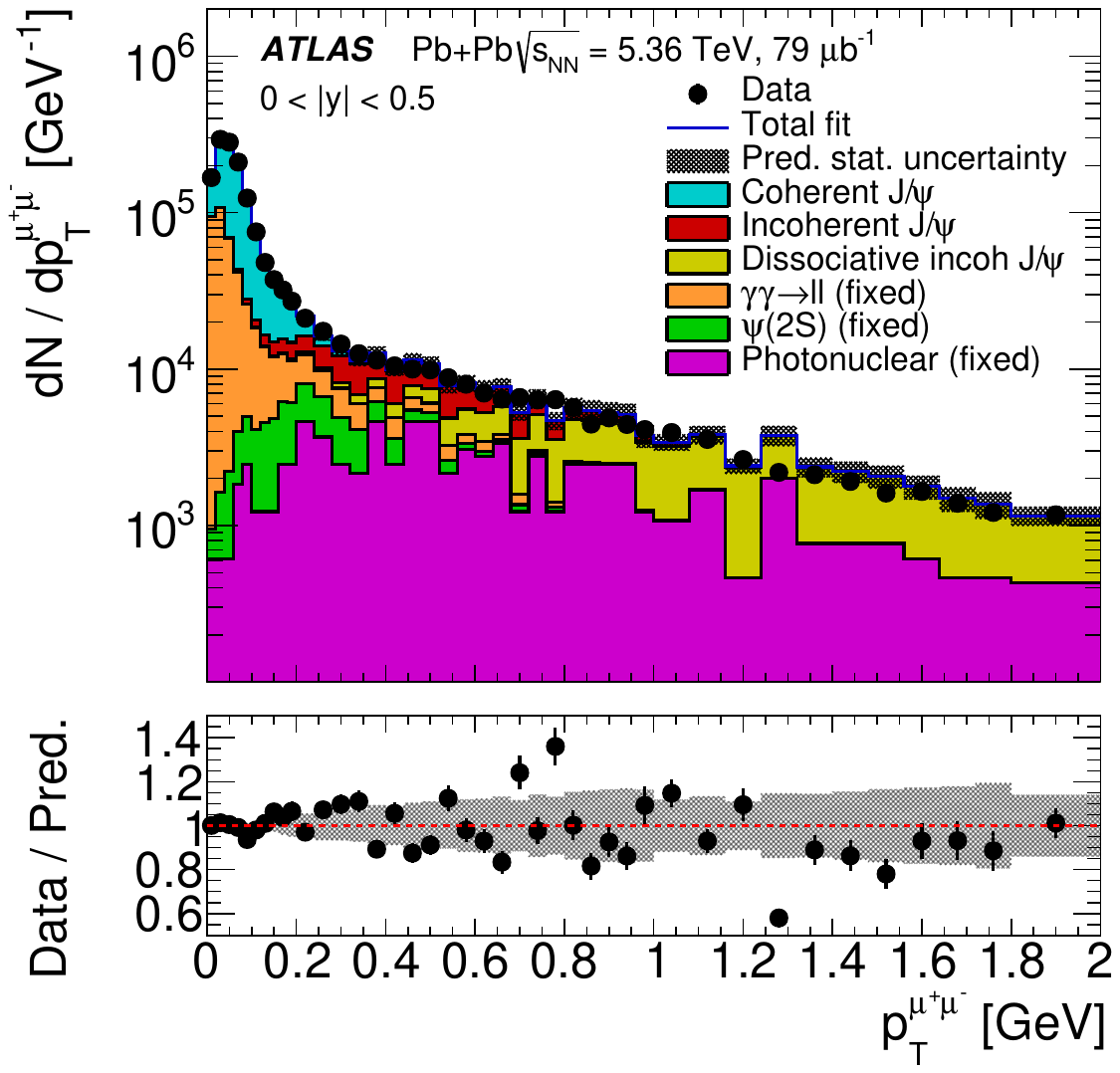}\vspace*{-0.5cm}}
\vspace*{-0.1cm}
\subcaptionbox{\label{fig:incoh_frac_fill2}}{\includegraphics[width=0.46\textwidth]{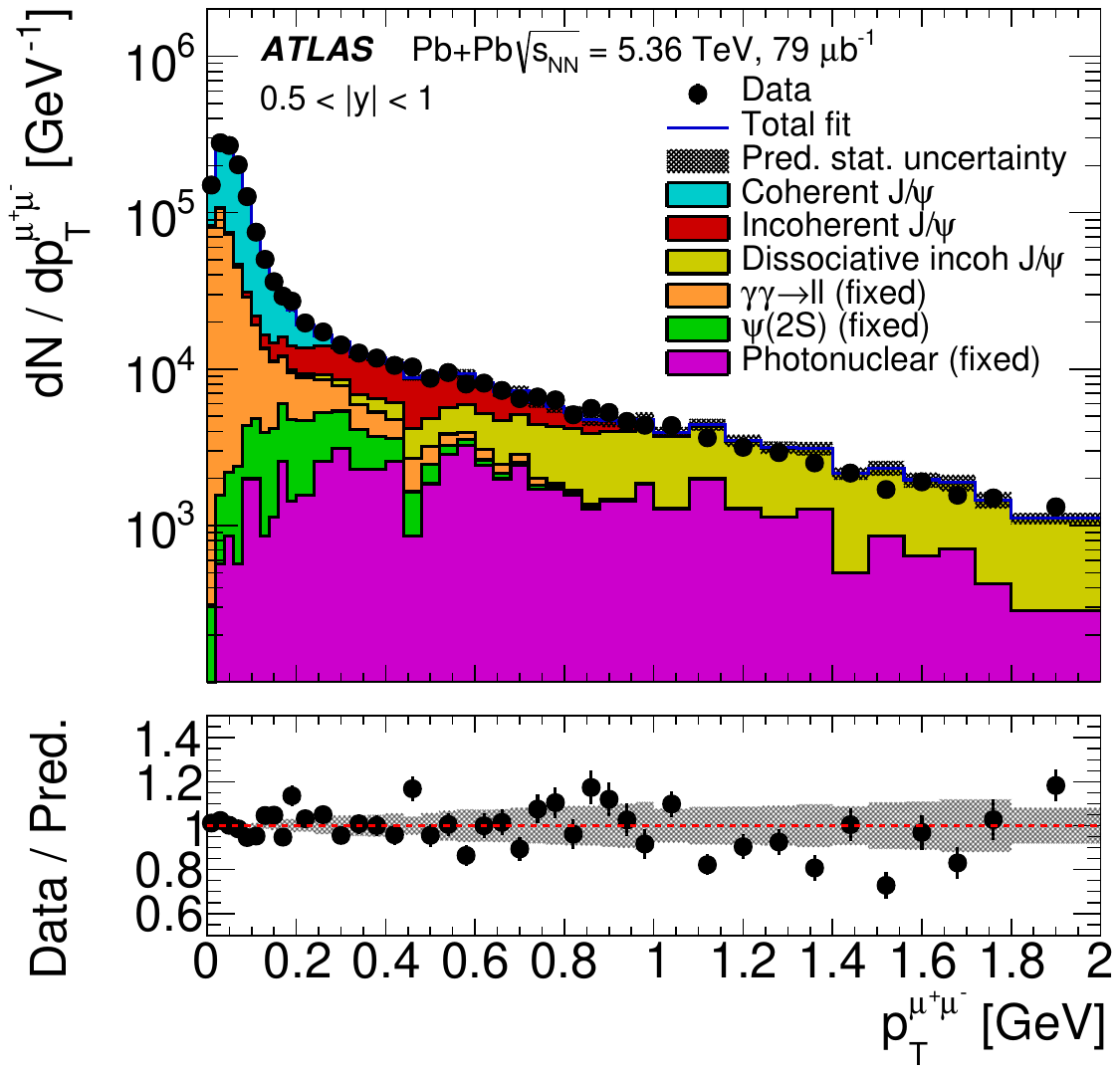}\vspace*{-0.5cm}}
\vspace*{-0.1cm}
\subcaptionbox{\label{fig:incoh_frac_fill3}}{\includegraphics[width=0.46\textwidth]{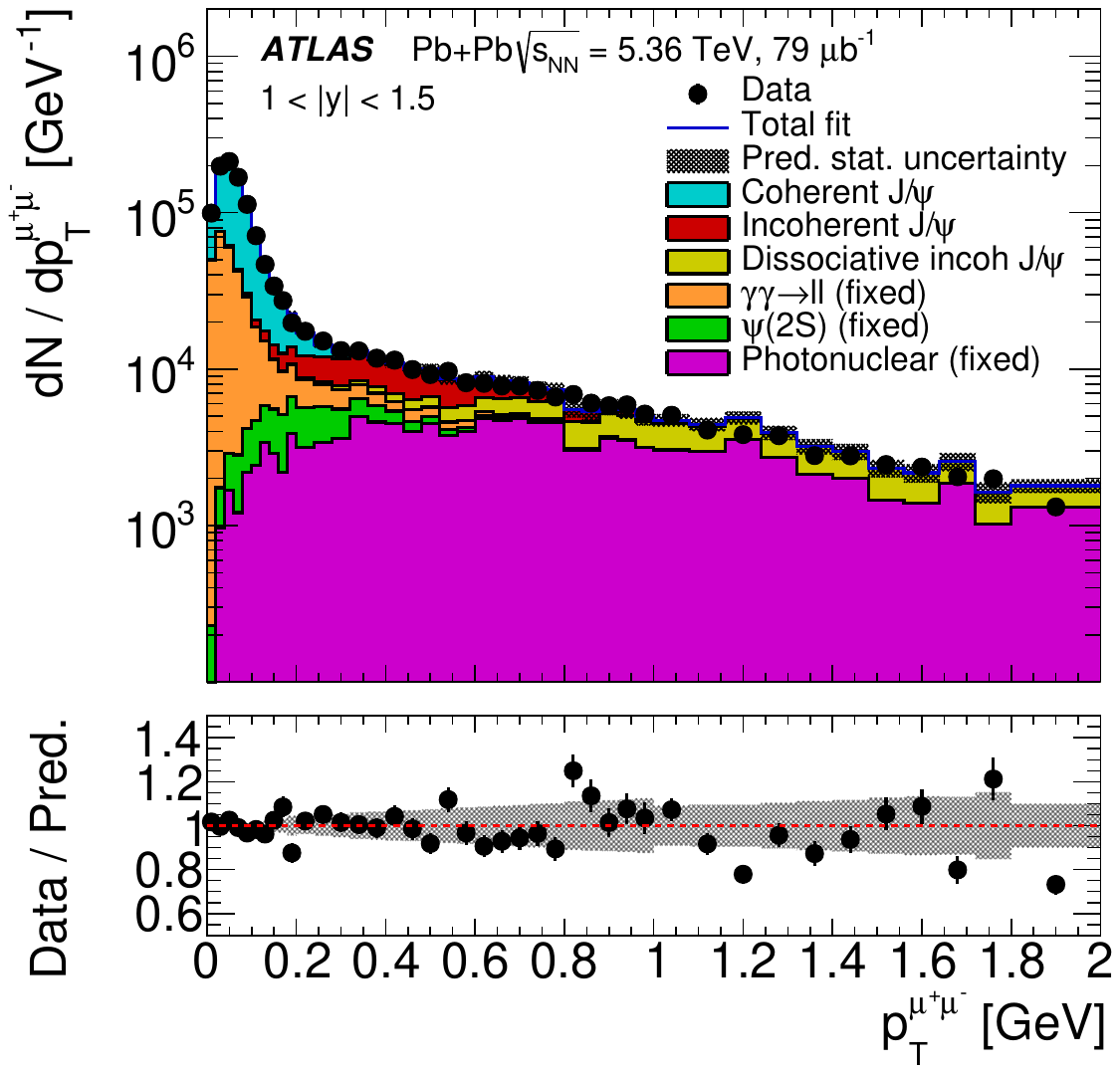}\vspace*{-0.5cm}}
\vspace*{-0.1cm}
\subcaptionbox{\label{fig:incoh_frac_fill4}}{\includegraphics[width=0.46\textwidth]{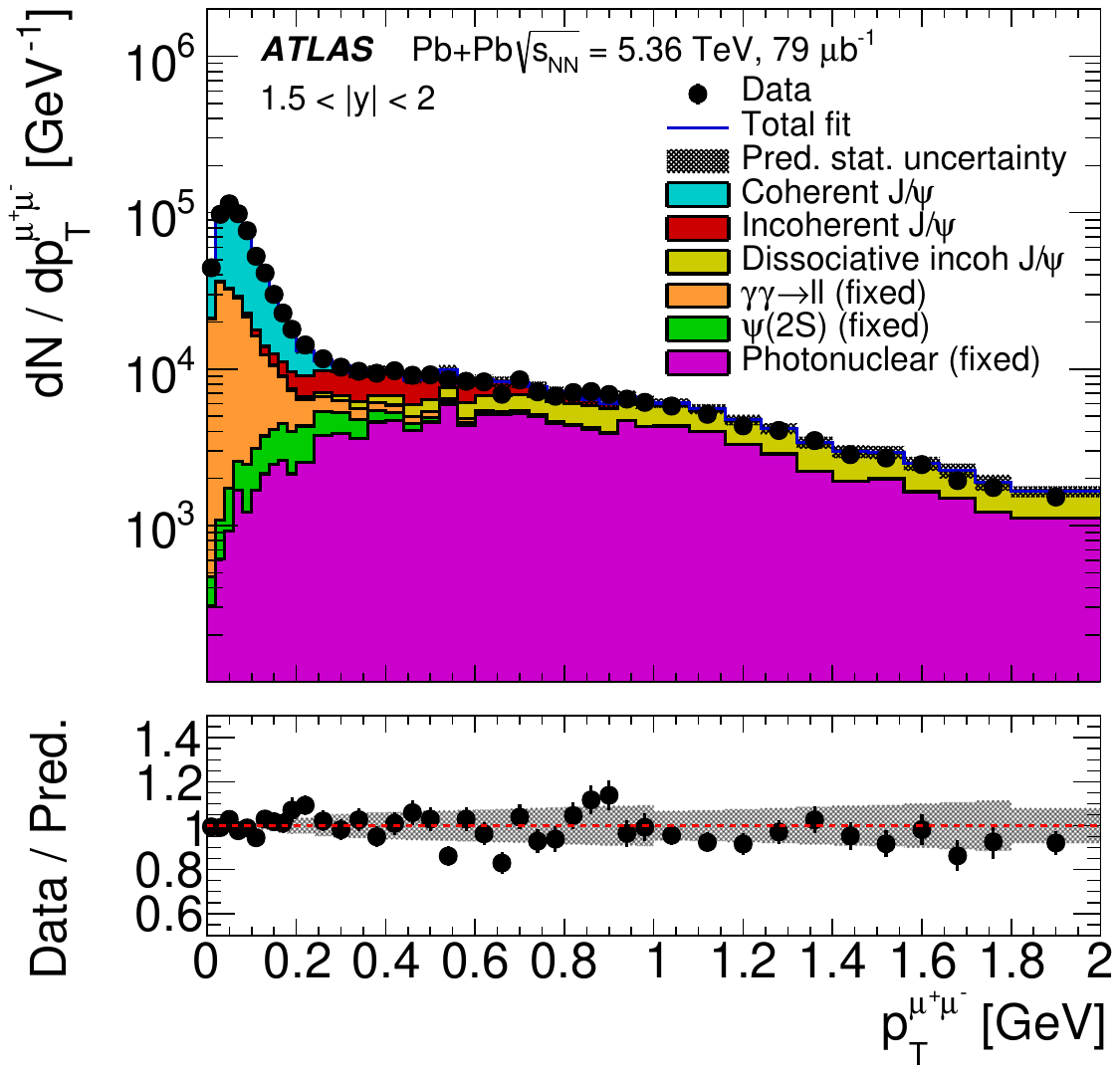}\vspace*{-0.5cm}}
\vspace*{-0.1cm}
\subcaptionbox{\label{fig:incoh_frac_fill5}}{\includegraphics[width=0.44\textwidth]{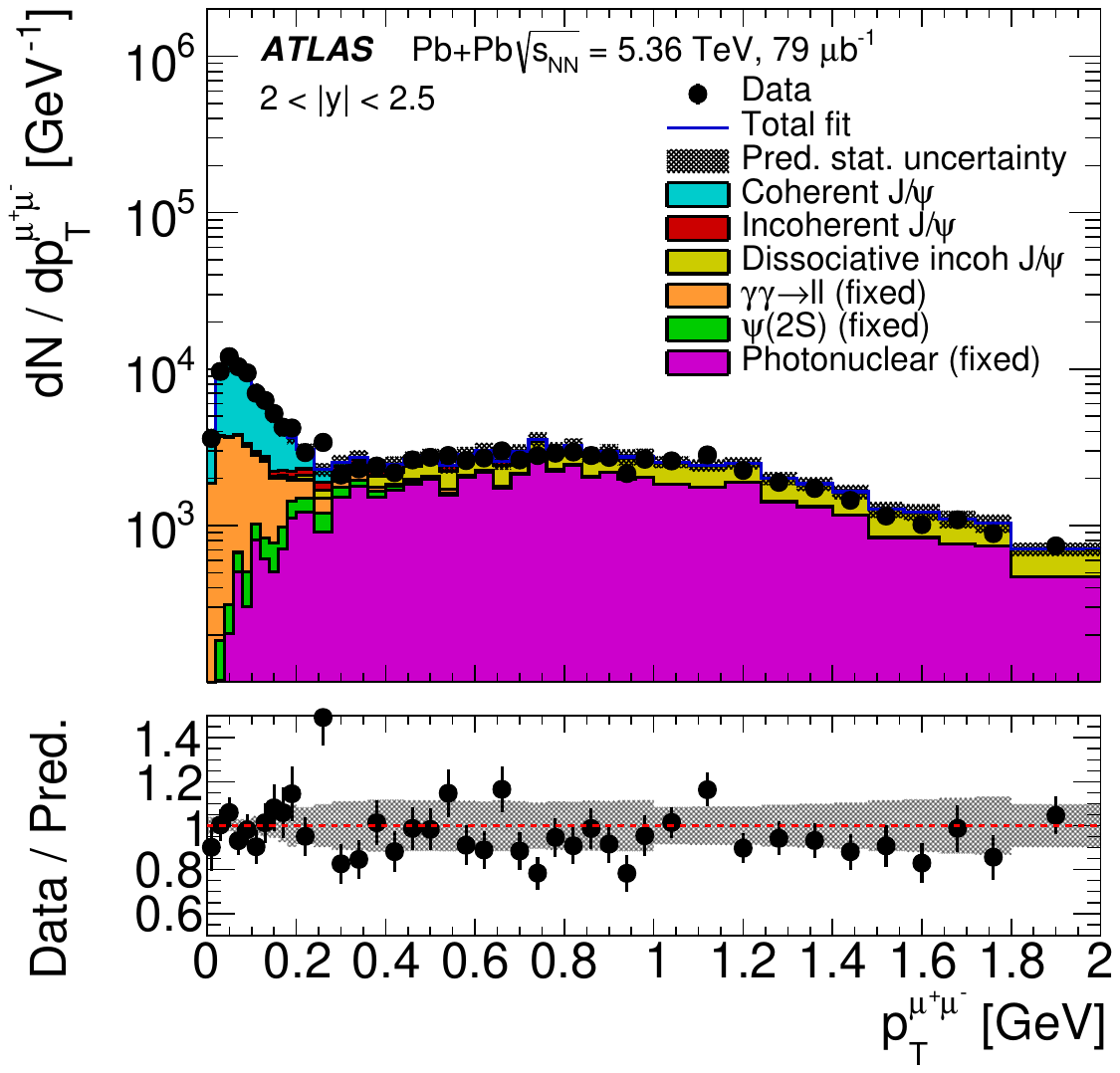}\vspace*{-0.5cm}}
\vspace*{-0.1cm}
\caption{ Demonstration of the coherent \jpsi signal yield
extraction. The exclusive two-track system \pT distributions in the
SR, after relaxing the $\pT^{\trktrk}<0.2$~GeV requirement, are fit
using templates from simulation, as described in the text.  The fits
are performed separately in (a--e) five $|y|$ intervals.  The total
fit result for each rapidity region is shown as a solid blue line.
Individual contributions to the fit are shown using different
colors. The bottom panel in each figure shows the ratio of data to
the total fit. For the calculation of $\pT^{\trktrk}$, the muon mass
hypothesis is used for each track. }
\label{fig:incoh_frac_fill}
\end{center}
\end{figure}

In these fits, the normalizations of the templates for $\psip$
feed-down and photonuclear pion pairs are kept fixed according to the
procedures explained in Sections~\ref{sec:psi2s} and \ref{sec:pions}.
The normalization of the dilepton continuum from the
$\gamma\gamma\to\ellell$ process is fixed by the results for the
background description of the two-track invariant mass fits.  In
addition, the $\ee/\mumu$ ratios for individual contributions are kept
fixed according to the results from the invariant mass fits.  Hence,
the only free parameters in the $\pT^{\trktrk}$ fits are the
normalization factors for coherent $\jpsi\to\ellell$, incoherent
$\jpsi\to\ellell$, and dissociative incoherent $\jpsi\to\ellell$ production.
For each rapidity interval, the binned maximum-likelihood fits are
performed for events in the $0< \pT^{\trktrk}<2$~GeV range.
Figure~\ref{fig:incoh_frac_fill} shows the $\pT^{\trktrk}$
distributions after the fit procedure.
The combined templates adequately describe the $\pT^{\trktrk}$ shapes observed in data.

While there is some overlap between events used in the invariant mass and $\pT^{\trktrk}$ fits, the impact of residual correlations is found to be negligible, as the incoherent
$\Jpsi$ contribution is small (a few percent in the signal region) and is primarily constrained by the high-$\pT^{\trktrk}$ tail, which lies outside the mass fit selection.

%
%
%
%
%
%
%
%
%
%
%
%
%
%
%
%
%
%
%
%
%
%


%

%


\section{Efficiency and acceptance corrections}
\label{sec:eff}

The coherent $J/\psi$ meson differential cross section is given by
\begin{equation}
\frac{\textrm{d}\sigma}{\textrm{d}y} = \frac{N^{\textrm{coh}}_{J/\psi\to\mumu}}{A\times \epsilon_C \times BR \times \mathcal{L}_{\textrm{int}}\times \Delta y}~,
\label{eq:xs}
\end{equation}
where $\Delta y$ denotes the rapidity interval,
$\mathcal{L}_{\textrm{int}}$ is the integrated luminosity of the data
sample, $BR$ is the $\jpsi\to\mumu$ branching fraction, $\epsilon_C$
and $A$ denote the efficiency and acceptance corrections, respectively, and
$N^{\textrm{coh}}_{\jpsi\to\mumu}$ is the raw yield of coherent
$\jpsi\to\mumu$ events extracted from the $\pT^{\trktrk}$ fits.  The
correction factor $\epsilon_C$ accounts for detector
inefficiencies, which can be factorized into reconstruction efficiency
and trigger efficiency components (the latter discussed in detail in
Section~\ref{sec:trigger_eff}).  It is defined as the ratio of the
number of generated events that satisfy the final selection criteria
after event reconstruction to the number of generated events within
the fiducial region.
The fiducial region is defined as two muons (after FSR) having
$\pT^{\mu}>1$~GeV, $|\eta^{\mu}|<2.5$, $2.9<m_{\mumu}<3.2$~GeV and
$\pT^{\mumu}<0.2$~GeV.  The factor $A$ is the acceptance for coherent
$\jpsi\to\mumu$ within the fiducial region. It is defined, using STARlight MC, as the fraction of
generated events in the given \jpsi rapidity interval and before the
emission of FSR from the decay muons (QED Born level) that satisfy the
fiducial requirements. This acceptance is determined entirely using the MC
signal sample and is used to extrapolate the measured cross section in
the fiducial region to the full phase space.
The migration between the individual $y$ bins is found to be below 0.5\%.

Table~\ref{tab:corrections} shows the numerical values of $\epsilon_C$ and $A$.
The values of $\epsilon_C$ vary between 64\% and 78\% for $|y|<2$.
For the $2<|y|<2.5$ rapidity bin, $\epsilon_C\approx 22\%$.  Similarly,
the values of $A$ are in the range of 51--63\% for $|y|<2$ region and
around 19\% for the $2<|y|<2.5$ region.
The acceptance loss comes mainly from the ID pseudorapidity coverage ($|\eta^{\mu}|<2.5$ requirement).

\begin{table}[]
\caption{Summary of efficiency and acceptance correction factors.}
\label{tab:corrections}
\centering
\begin{tabular}{c|c|c}
\hline
Rapidity range & $A$ & $\epsilon_C$ \\
\hline
$|y|<0.5$   & 0.633  $\pm$ 0.002 (stat.)  & 0.722 $\pm$ 0.002 (stat.) $\pm$ 0.016 (syst.) \\
$0.5<|y|<1$ & 0.632  $\pm$ 0.002 (stat.)  & 0.776 $\pm$ 0.002 (stat.) $\pm$ 0.013 (syst.) \\
$1<|y|<1.5$ & 0.633  $\pm$ 0.002 (stat.)  & 0.720 $\pm$ 0.002 (stat.) $\pm$ 0.020 (syst.) \\
$1.5<|y|<2$ & 0.510  $\pm$ 0.002 (stat.)  & 0.643 $\pm$ 0.003 (stat.) $\pm$ 0.020 (syst.) \\
$2<|y|<2.5$ & 0.186 $\pm$ 0.001 (stat.) & 0.227 $\pm$ 0.002 (stat.) $\pm$ 0.019 (syst.) \\
\hline
\end{tabular}
\end{table}


%

%

\section{Trigger efficiency measurement}
\label{sec:trigger_eff}

The efficiency of the L1 TRT trigger is determined using an
independent set of triggers based on a ZDC selection at L1.  The
supporting ZDC trigger selects events having ZDC activity consistent
with the presence of at least one neutron on either side and none
on the opposite side, accompanied by a total maximum
transverse energy of 200~GeV.  At the HLT, a 5~GeV maximum transverse
energy in each FCal region, and at least one track in the ID with
$\pT>200$~MeV, are required.  These triggers are sensitive to
coherent $\jpsi\to\mumu$ events with single electromagnetic
dissociation of one of the outgoing ions~\cite{ALICE:2023jgu}.  The
ZDC triggers are fully independent of the ID, and thus serve as an
unbiased reference for the L1 TRT trigger.

The offline event selections for the L1 TRT trigger efficiency evaluation are
similar to that used in the actual $\jpsi\to\mumu$ analysis.
To extract the L1 TRT trigger efficiency for muon pairs from
$\jpsi\to\mumu$ decay, fits to the 2--track invariant mass
distributions are performed, similar to the nominal signal
analysis (Section~\ref{sec:mass_fit}).

The L1 TRT trigger efficiency has a significant
dependence on the track $|\eta|$, reflecting the physical structure of the TRT\@.
To take this into account, the
efficiency is derived in two dimensions, in bins of
$|\eta_{\textrm{1}}|$ and $|\eta_{\textrm{2}}|$, where
$|\eta_{\textrm{1}}|$ ($|\eta_{\textrm{2}}|$) denotes the absolute
pseudorapidity of a track with lower (higher) $|\eta|$ value.  The
efficiency is studied in six $|\eta|$ regions:
$|\eta|<0.7$, corresponding to the TRT barrel region;
$0.7\leq|\eta|<0.9$ and $0.9\leq|\eta|<1.1$, corresponding to the transition region;
$1.1\leq|\eta|<1.8$, corresponding to the TRT endcap regions; and
$1.8\leq|\eta|<2$ and $2\leq|\eta|<2.5$, corresponding to the TRT
endcap edge.

\begin{figure}[!t]
\begin{center}
\subcaptionbox{\label{fig:trig1}}{\includegraphics[width=0.49\textwidth]{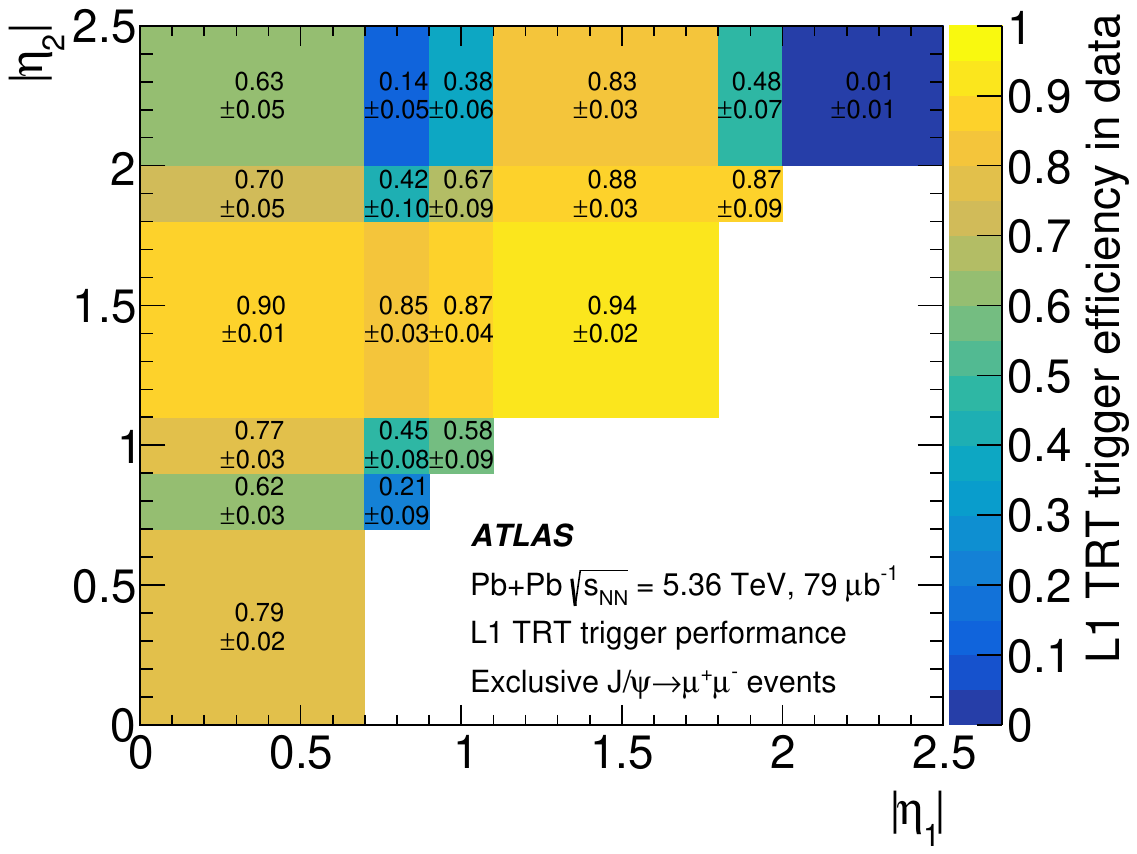}}
\subcaptionbox{\label{fig:trig2}}{\includegraphics[width=0.49\textwidth]{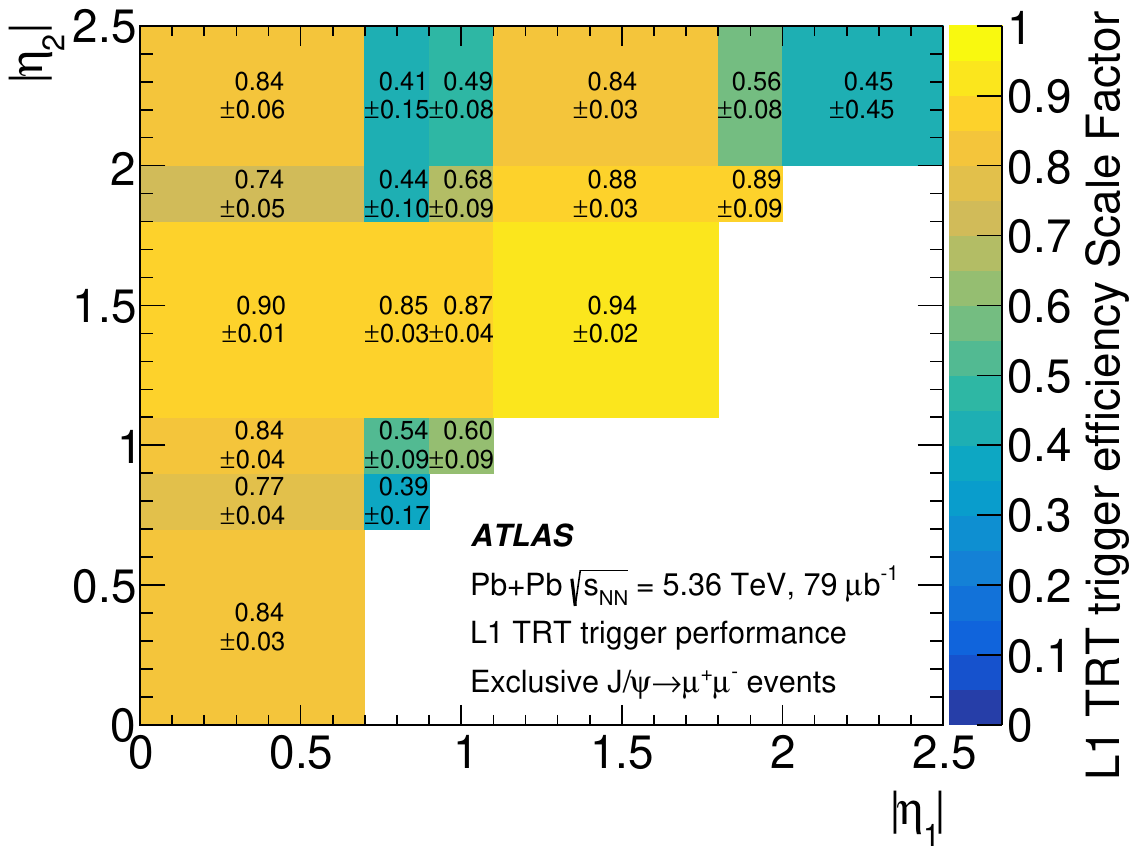}}

\caption{
L1 TRT trigger (a) efficiency and (b) scale factors for dimuon events in data as a function of $|\eta_{\textrm{1}}|$ and $|\eta_{\textrm{2}}|$.
Numerical values of trigger efficiency and scale factors are also shown, along with the respective statistical uncertainties.
}
\label{fig:trigger_eff}
\end{center}
\end{figure}

The $\jpsi\to\mumu$ yields are extracted from individual
mass fits in each $(|\eta_{\textrm{1}}|,~|\eta_{\textrm{2}}|)$
bin, separately for events satisfying the supporting trigger and for
events satisfying both the supporting and L1 TRT triggers.  The resulting
yields are then used to calculate the L1 TRT trigger efficiencies.
These are then compared with the efficiencies calculated similarly in signal MC
to derive per-event trigger efficiency scale factors (SFs).
Figure~\ref{fig:trigger_eff} shows the measured L1 TRT trigger
efficiencies and SFs for dimuon events as a function of
$|\eta_{\textrm{1}}|$ and $|\eta_{\textrm{2}}|$.  The observed
efficiency is around 60--90\% if at least one muon is found in the TRT
barrel region, and is approximately 80--90\% if at least one muon is in
the TRT endcap region.  The efficiency drops typically to 14--50\% if
there is a muon in either transition or TRT endcap edge regions.
Similarly, the extracted SF values vary with $|\eta_{\textrm{1}}|$
and $|\eta_{\textrm{2}}|$, as seen in Figure~\ref{fig:trig2}.
The mismodeling of the L1 TRT trigger efficiency in MC simulation observed in Figure~\ref{fig:trig2} is due to the TRT module-level deadtime that is not accounted for in the simulation.

The measured SFs are used to correct the simulated signal and
background MC events.
They are applied as multiplicative factors to the MC event weights.
In particular, they improve the $\epsilon_C$ factors used in Eq.~\ref{eq:xs} by providing an accurate description of the L1 TRT trigger efficiency as a function of $y$.
Due to the limited size of the data sample from the
supporting trigger, the same SFs (as derived for muon pairs) are
applied also to the dielectron final states.  Possible differences
between the SFs for muons and electrons are taken into account in the
signal mass fit procedure (Section~\ref{sec:mass_fit}), by allowing the
relative yields of $\jpsi\to\ee$ and $\jpsi\to\mumu$ events to float.
No dependence of the L1 TRT
trigger SFs on other aspects of the muon-pair kinematics,
including the individual $\pT$ of the muons, is observed.
Similarly, the muon charge and azimuthal angle efficiency-dependence is well modelled in the MC simulation.

The HLT selection efficiency of the signal trigger is studied using
exclusive 2-track events satisfying the supporting trigger with the same
L1 selection as the signal trigger, but with relaxed HLT
requirements. The same event selection as the SR is also used.  The HLT
selection efficiency in data is found to be above 99\% for the signal
process.  Good agreement is found between the HLT efficiency in data
and in MC simulation, and thus no additional HLT correction is applied
to MC-simulated events.


%

%
%
%

%


\section{Systematic uncertainties}
\label{sec:Sys}

Systematic uncertainties in the coherent \jpsi cross section measurements arise from the reconstruction of
ID tracks, the trigger efficiency estimation, signal and background modeling, the integrated luminosity uncertainty, and the $\jpsi\to\mumu$ branching fraction uncertainty.

\subsection{Detector material modeling}

For evaluating the systematic uncertainties of the ID material
description, alternative MC-simulated signal and background samples
are produced.  These implement variations in the amount of material in different
ID regions used in the detector simulation:
the passive material of the ID is scaled up by 5\%, the passive material of the
innermost ID layer is scaled up by 10\%, or the passive material in
the services region is scaled up by 25\%.  These variations capture
the full range of data--MC differences observed in studies of the ID
material~\cite{PERF-2015-07}.
The impact of these variations on the analysis can be two-fold:
a change in the signal efficiency due to changes in track reconstruction efficiency, and
changes in the shape of the $m_{\trktrk}$ distribution ($m_{\trktrk}$ lineshape).
The differences between the nominal signal yield and the one measured using
each material variation is taken as systematic uncertainty.  Then the
symmetrized contributions are added in quadrature to calculate the total
systematic uncertainty.

The resulting variations for the signal efficiency are typically
between few per mil ($0<|y|<2.0$) and 1.6\% ($2.0<|y|<2.5$).
The variations resulting from different template shape changes are summarized in
Table~\ref{tab:sys_shapes}.
This table shows how shape variations of individual distributions ($\jpsi\to\mumu$ invariant mass,
$\jpsi\to\ee$ invariant mass and $\pT^{\trktrk}$) influence the measured yields.
The total uncertainty accounting for
correlated variations in all three distributions is also shown.
\begin{table}[t!]
\centering
\caption{Systematic uncertainty in the measured coherent $J/\psi\to
\mumu$ yields due to the impact of variations in detector geometry description on $m_{\trktrk}$ and $\pT^{\trktrk}$ template shapes.  The last row shows the total uncertainty
accounting for a correlated variation of all template shapes.}
\begin{tabular}{l| c c c c c }
\hline
Absolute rapidity interval & 0--0.5     & 0.5--1     & 1--1.5     & 1.5--2     & 2--2.5     \\
\hline
$J/\psi\to \mumu$ $m_{\trktrk}$ lineshape & 1.4\% & 1.2\% & 0.7\% & 0.7\% & 0.6\%  \\
$J/\psi\to \ee$ $m_{\trktrk}$ lineshape     & 0.7\% & 0.6\% & 0.3\% & 0.2\% & 0.1\%  \\
$\pT^{\trktrk}$ shape   & 0.1\% & 0.1\% & 0.1\% & 0.2\% & 0.3\%  \\
\hline
Total ID material unc. on signal yield     & 2.1\% & 1.8\% & 0.9\% & 1.0\% & 0.8\%  \\
\hline
\end{tabular}
\label{tab:sys_shapes}
\end{table}
Although the electron-track momentum resolution is expected to be more
sensitive to material variations than the muon-track momentum
resolution, the leading contribution to the signal yield uncertainty
is the $J/\psi\to \mumu$ mass lineshape.  The shape variations of the
$p_{\textrm{T}}$ spectrum give negligible impact on the final results.

\subsection{Trigger efficiency corrections}

The statistical uncertainties in the L1 TRT trigger SFs are
propagated using pseudo-experiments in which the SF values are
randomly shifted %
according to the mean and standard deviation of the correction factor in each bin.
This uncertainty affects the measured
cross sections by about 1.5--2\% in the $0<|y|<2.0$ range, and it
grows to 7\% in the $2.0<|y|<2.5$ region.

The systematic uncertainties in the L1 TRT trigger SFs are estimated by varying the selection criteria described in Section~\ref{sec:trigger_eff}, including tightening the $|d_0|$ requirement for each track and loosening the $\pT^{\trktrk}$ requirement.
They are then propagated in the nominal analysis by the offset method in which the SF values are coherently
shifted upwards and downwards by one standard deviation and the
magnitude of the change in the measurement is computed.  These
variations result in an 1--3.5\% change of the measured cross sections.

\subsection{Signal and background modeling}

The uncertainty in the signal modeling used in the signal extraction
procedure ($\pT^{\trktrk}$ fits) is estimated by taking the full
difference between the fits using the signal $\pT^{J/\psi}$ shape
reweighted to LHCb data and the signal $\pT^{J/\psi}$
shape from STARlight.
This results in 0.3--1.6\%
variations in the signal yield, which is taken as
systematic uncertainty.
The effect of the signal $\pT^{J/\psi}$ shape uncertainty in $\epsilon_C$ and $A$ is found to be negligible.

To evaluate the impact of the mass fit methodology assumptions, an alternative fit is performed in which no additional scaling is applied between the
$J/\psi\to\ee$ and $J/\psi\to \mumu$ components beyond what is already present in the MC simulation. The difference between this alternative and the nominal fit result is taken as an additional systematic uncertainty.
The difference typically amounts to 1--3.7\% as a function of $y$.

The uncertainties in the modeling of the $\gamma\gamma\to\ellell$ contribution are
estimated by using, for the alternative mass shape parameterisation, a second order polynomial
instead of an exponential function.
This results in a 0.4--3.2\% impact on the measured cross sections.

To assign an uncertainty due to the incoherent $\jpsi$ background modeling,
the $\pT^{\trktrk}$ fits are repeated for different fit ranges
in terms of the maximum $\pT^{\trktrk}$ value, by varying its value between 1~GeV and 3~GeV.
This typically results in a 0.2--0.3\% impact on the measured cross sections.

A 20\% relative uncertainty in the $\psi(2S)$ feed-down
background contribution is assigned to cover data-to-simulation differences observed in the $\psi(2S)$ CR.
This results in an approximately 0.5\% impact on the measured cross sections.

\subsection{Other systematic uncertainties}
\label{sec:sysother}

The track veto applied in the event selection could be biased by the presence of additional UPC interactions in the same or separate Pb+Pb collision, resulting in the production of extra charged particle pairs, for example from the  $\gamma\gamma\to\ee$ process or from the coherent $\rho^0\to\pi^+\pi^-$ production (see Section~\ref{sec:Results}).
To check the impact of such processes on the results, the analysis is repeated with a relaxed track veto condition, to allow events with up to four reconstructed tracks instead of exactly two.
The difference between the extracted signal yield from the modified (2--4 tracks) and nominal (exactly two tracks) track-veto requirements is taken as a systematic uncertainty. This difference is 2.1\% with no dependence on $y$.

To cross-check event activity in the SR beyond the ID
acceptance ($|\eta|<2.5$), topological clusters of calorimeter-cell
energy deposits~\cite{PERF-2014-07} are studied. These topoclusters,
calibrated to account for noise suppression, are required to have
transverse energy $E_\mathrm{T} > 0.2$~\GeV and pseudorapidity in the
range of $3.1 < |\eta| < 4.9$. Additionally, they must satisfy the
cell significance criteria for the measured energy, as outlined in
Ref.~\cite{STDM-2011-01}, to suppress contributions from electronic
noise fluctuations.  Approximately 4\% of events
in the SR contain at least one additional topocluster. This
observation is consistent with expectations from incoherent
backgrounds and the production of extra particle pairs in the forward
region for the signal process. Consequently, no additional systematic
uncertainty is assigned.
This cross-check also ensures that potential backgrounds with extra hadronic activity, such as those from peripheral coherent $J/\psi$ production or from beam-gas interactions, are essentially negligible.

The uncertainty in the integrated luminosity of the data sample is
2.7\%.  It is derived from the calibration of the luminosity following
a methodology similar to that detailed in Ref.~\cite{DAPR-2021-01},
using the LUCID-2 detector~\cite{LUCID2} for the baseline
luminosity measurements.

\begin{figure}[b]
\begin{center}
\includegraphics[width=0.6\textwidth]{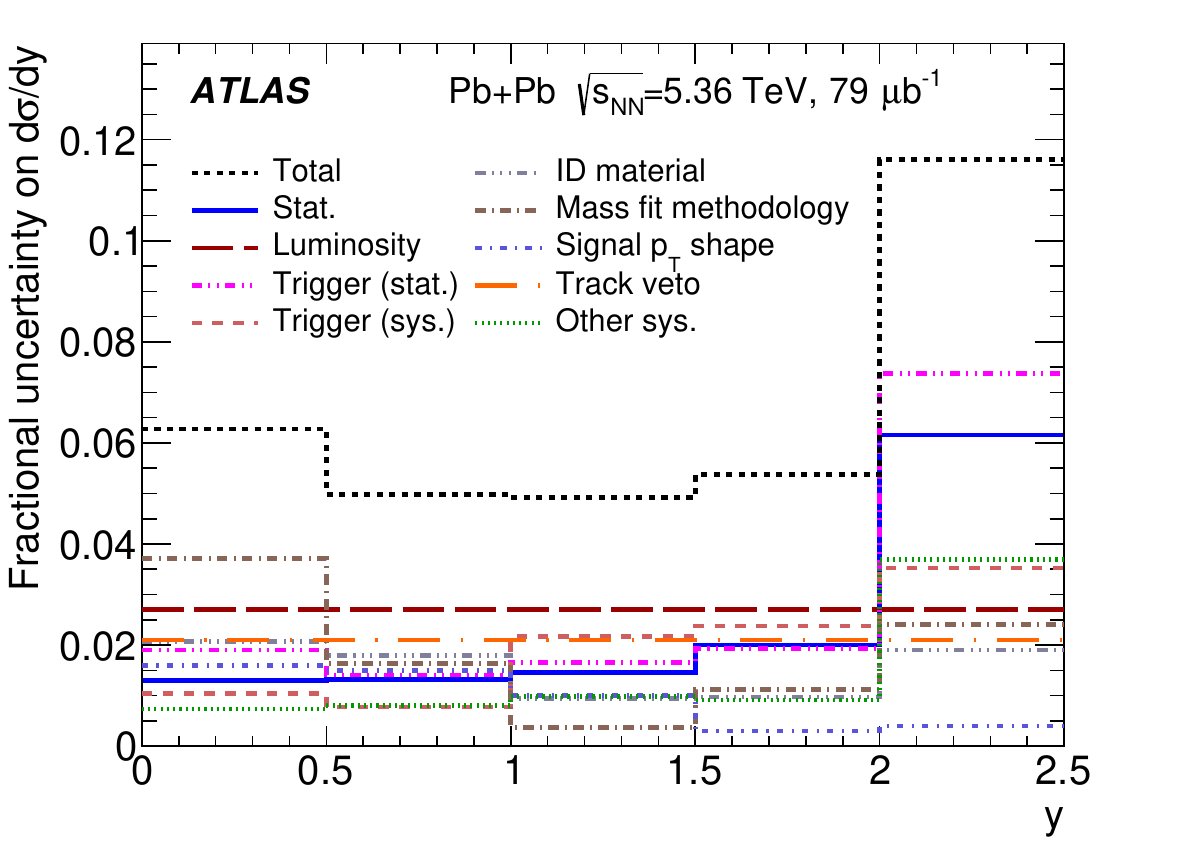}
\caption{
Breakdown of the individual uncertainty contributions as a function of $y$. The total measurement uncertainty, calculated by adding all contributions in quadrature, is also shown.
}
\label{fig:systot}
\end{center}
\end{figure}

The relative uncertainty in the $J/\psi\to \mumu$ branching ratio is 0.5\%
~\cite{ParticleDataGroup:2024cfk}.
The total measurement uncertainty includes both the statistical and
systematic components added in quadrature.  Figure~\ref{fig:systot}
shows the breakdown of the uncertainties for the differential cross
section measurement.  The total measurement uncertainty is around
5--6\%, except the last rapidity bin where the statistical uncertainty
and the trigger scale factor statistical uncertainties are larger.

%

%


\section{Results}
\label{sec:Results}

\begin{figure}[t!]
\begin{center}
\subcaptionbox{\label{fig:xs_meas1}}{\includegraphics[width=0.49\textwidth]{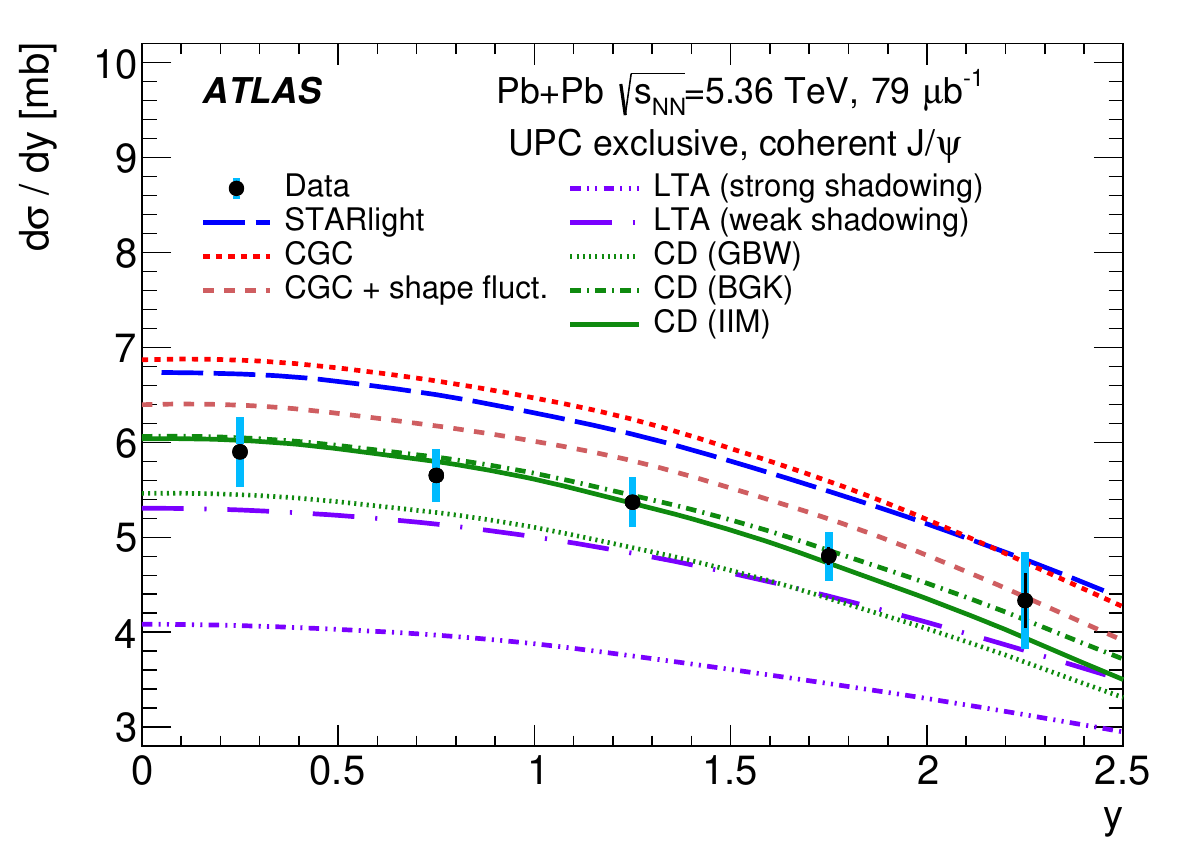}}
\subcaptionbox{\label{fig:xs_meas2}}{\includegraphics[width=0.49\textwidth]{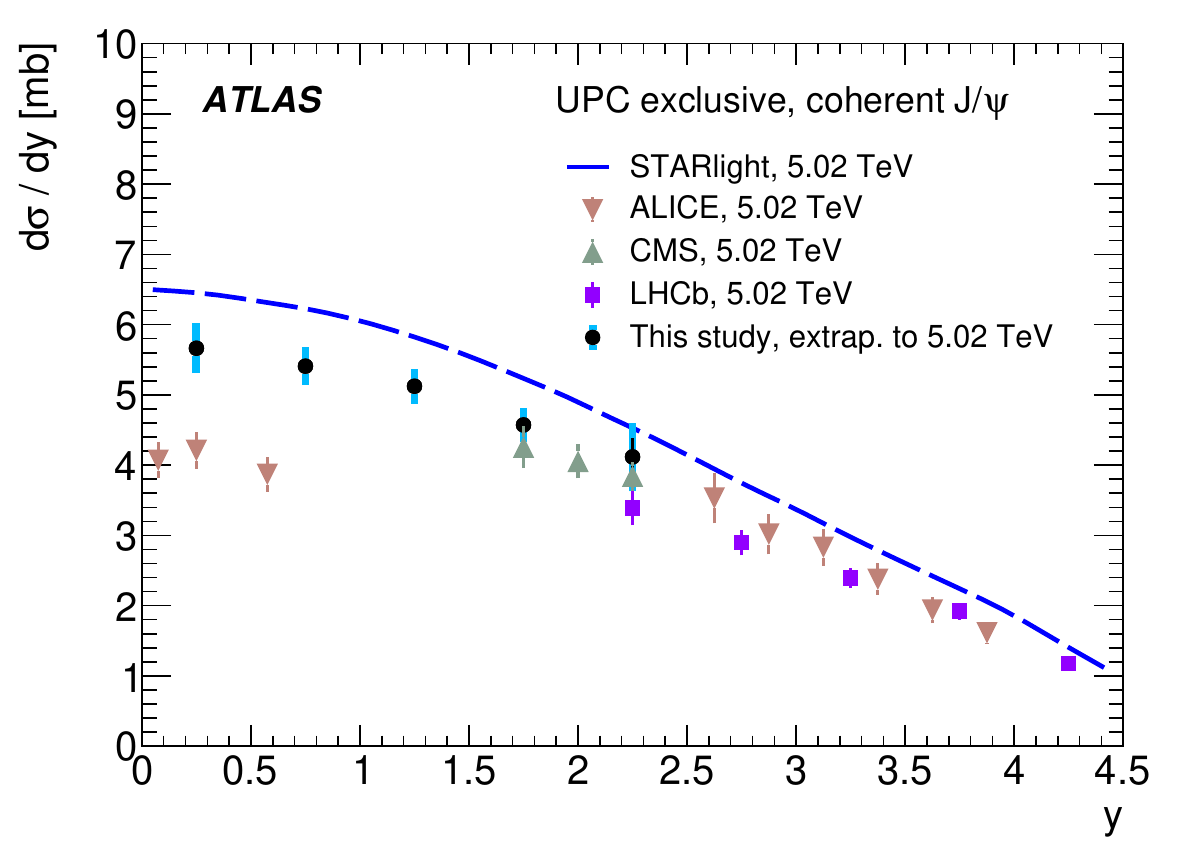}}
\caption{ (a) Measured differential cross section for coherent exclusive
$\jpsi$ meson production in Pb+Pb UPC at $\sqrt{s_{\textrm{NN}}}=5.36$~TeV as a function of
$\jpsi$ rapidity (averaged over both sides). Data (points) are compared with several model
predictions: STARlight 2.0 (using classical Glauber option), two versions of a color glass
condensate (CGC) model~\cite{Mantysaari:2022sux,
Mantysaari:2023xcu}, color dipole (CD) models (BGK, GBW and IIM)~\cite{Luszczak:2019vdc}, and leading-twist approximation (LTA) weak
and strong shadowing models~\cite{Guzey:2016piu}.  For each measurement, the thin vertical bar represents
the statistical uncertainty, and the thick bar represents the total
uncertainty (with statistical and systematic uncertainties added in quadrature).
(b) Measured differential cross section for
coherent $\jpsi$ meson production in Pb+Pb UPC at $\sqrt{s_{\textrm{NN}}}=5.02$~TeV as a
function of $\jpsi$ rapidity. The plot displays the previous
results from the ALICE~\cite{ALICE:2019tqa, ALICE:2021gpt},
CMS~\cite{CMS-HIN-22-002} and LHCb~\cite{LHCb:2022ahs} experiments,
together with the results presented here after extrapolation to 5.02~TeV.
Data (points) are compared with STARlight predictions.
}
\label{fig:xs_meas}
\end{center}
\end{figure}

The coherent exclusive $\jpsi$ meson differential cross sections extrapolated to the full phase space are
obtained using Eq.~\ref{eq:xs} and include the full set of
statistical and systematic uncertainties.  The measured differential
cross sections are shown in Figure~\ref{fig:xs_meas1}.  Measured cross
section values are compared with predictions from STARlight (using classical Glauber option) and two
calculations based on the color glass condensate (CGC) approach, one
of which also includes the effect of nucleon shape
fluctuations~\cite{Mantysaari:2022sux, Mantysaari:2023xcu}. Both STARlight and CGC
models use HERA $\gamma p\to J/\psi p$ data as an input.
Predictions from other theoretical calculations are also shown in Figure~\ref{fig:xs_meas1} for comparison.
The color dipole (CD) models~\cite{Luszczak:2019vdc}, with different parameterizations (labeled BGK, GBW, and IIM), assume quark--antiquark dipole scattering off the nuclear target.
The leading-twist approximation (LTA)~\cite{Guzey:2016piu} is a perturbative QCD calculation that accounts for nuclear shadowing effects arising from multi-nucleon interference. Both weak and strong shadowing scenarios are shown.
Among the models considered, the color dipole calculations provide the best description of the data.
The data also agree with NLO perturbative QCD calculations from Ref.~\cite{Eskola:2022vpi} (not shown in Figure~\ref{fig:xs_meas1}), albeit within large scale uncertainties.

To quantify the level of nuclear suppression observed at mid-rapidity, the results obtained for $|y|<0.5$ are compared with the impulse approximation (IA) model~\cite{Guzey:2013jaa} as implemented in STARlight. This model is based on a simple scaling of the experimental data from exclusive $J/\psi$ photoproduction off protons, includes the nuclear form factor, and neglects all other nuclear effects except for coherence. The ratio of the measured cross section to the IA model prediction is found to be $S_{\textrm{Pb}}^2(|y|<0.5)=0.57 \pm 0.04$, indicating significant nuclear suppression.

To compare this measurement with previous measurements from the
ALICE, CMS and LHCb Collaborations at $\sqrt{s_{\textrm{NN}}}=5.02$~TeV, the measured cross
sections need to be extrapolated from 5.36~TeV to 5.02~TeV.
The extrapolation factor is estimated by using STARlight and
is found to reduce the cross sections by
4--5\%, depending on the rapidity.\footnote{A similar extrapolation factor, of the order of 4--5\%, is obtained when using the CGC-based and CD-based models.}
Figure~\ref{fig:xs_meas2} presents
the results from the ALICE~\cite{ALICE:2019tqa, ALICE:2021gpt},
CMS~\cite{CMS-HIN-22-002} and LHCb~\cite{LHCb:2022ahs} experiments,
together with the results presented here, after extrapolation to 5.02 TeV.
While the extrapolated measurement matches quite well (within 1--1.5 standard deviation)
previously measured values at forward rapidities ($1.5<y<2.5$),
the extrapolated results at mid-rapidity ($0<y<1$) differ
significantly from the existing ALICE measurements~\cite{ALICE:2021gpt}.

A significant difference between the extrapolated measurement
and the previous ALICE result may be due to the production of
additional particle pairs accompanying the produced $\jpsi$ meson,
which are rejected by selections on the ALICE forward counters~\cite{ALICE:2021gpt}.
The production rate of these extra particle pairs is dominated by soft
$\ee$ pair production.  However, at higher pair
invariant masses, the production of $\mumu$ pairs and $\pi^+\pi^-$
pairs from coherent $\rho^0$ decays is also possible~\cite{Klein:1999qj,Baur:2003ar}.
For the
measurements of exclusive processes, these extra pairs may violate the
exclusivity requirements, as suggested in
Ref.~\cite{Harland-Lang:2021ysd}, and induce the rejection of signal
events.
As shown in Section~\ref{sec:sysother}, this measurement has little sensitivity to additional particles created in the same UPC collision,
presumably due to the extra particles being typically emitted outside the
$\eta$ and $\pT$ acceptance of the ID.

\begin{figure}[b!]
\begin{center}
\subcaptionbox{\label{fig:jpsi_rho1}}{\includegraphics[width=0.49\textwidth]{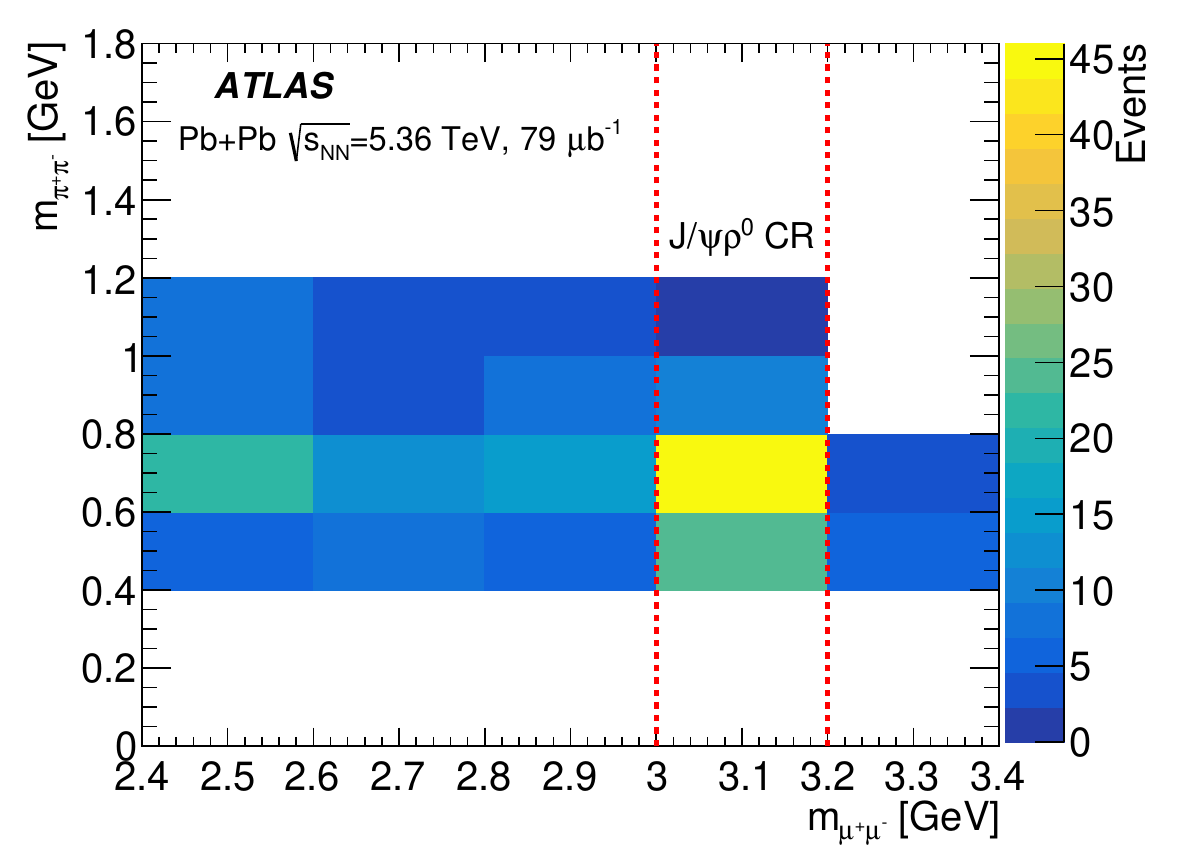}}
\subcaptionbox{\label{fig:jpsi_rho2}}{\includegraphics[width=0.49\textwidth]{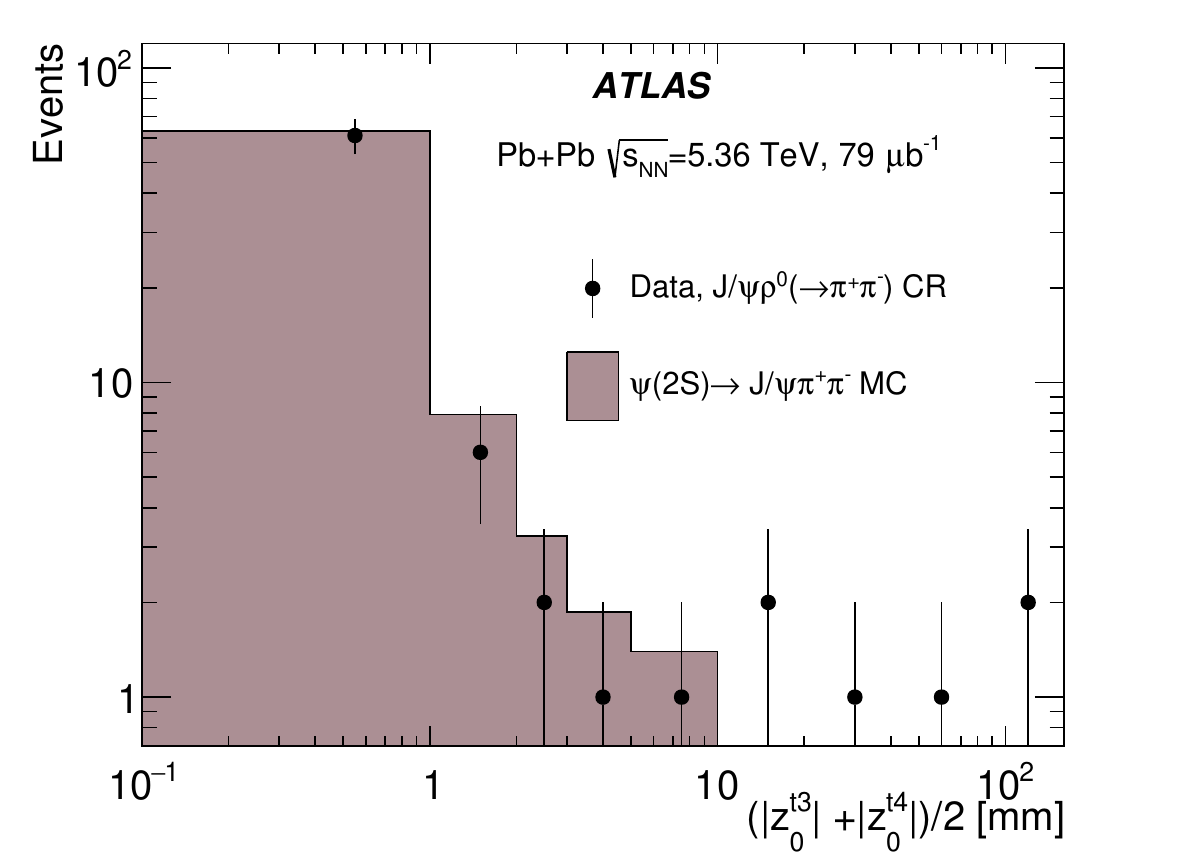}}
\caption{
(a) Correlation between the invariant mass of the two leading tracks and the invariant mass of the two sub-leading tracks for events satisfying $J/\psi \rho^0$ selection, after relaxing the $3<m_{\trktrk}<3.2$~GeV requirement. The boundaries at $m_{\pi^+\pi^-}=0.4$~GeV and $m_{\pi^+\pi^-}=1.2$~GeV are due to selection requirements correlated with $m_{\pi^+\pi^-}$.
(b) Average longitudinal impact parameter distribution of the third and fourth leading track for the $J/\psi \rho^0$ selected events in data (points). The distribution is compared with simulated $\psip \to\jpsi(\to\ellell)\pi^+\pi^-$ MC events.}
\label{fig:jpsi_rho}
\end{center}
\end{figure}

Although ATLAS cannot study forward $\ee$ production associated with the $\jpsi$,
primarily due its limited acceptance in high pseudorapidity region for low energy electrons, a related check has been performed
using the $J/\psi(\to\ellell)\rho^0(\to \pi^+\pi^-)$ process.
For example, the relative rate of coincident production of $J/\psi$ and $\rho^0$,
compared with
those from separate collisions, can be checked.
Figure~\ref{fig:jpsi_rho1}
presents the correlation between $m_{\trktrk}$ and
$m_{\pi^+\pi^-}$ for events satisfying the $J/\psi \rho^0$ control region selections summarized in Section~\ref{sec:psi2s}, where a clear simultaneous $J/\psi \rho^0$ contribution is visible.
Because the $\rho^0$  is a broad resonance, no further selection is applied to the  $m_{\pi^+\pi^-}$ spectrum.

To study the contributions of $J/\psi \rho^0$ events coming from the
same or separate Pb+Pb collisions, it is examined whether the two
particle pairs come from the same vertex or from different ones.
Since the low $\pT$ of the $\rho^0$ decay tracks often leads to no
vertex being reconstructed, this is done by checking the probability
of the $\rho^0$ tracks to have a large longitudinal impact parameter
relative to the reconstructed $\jpsi$ vertex, $|z_0|>2$~mm.
Figure~\ref{fig:jpsi_rho2} shows the average longitudinal impact
parameter distribution of the third and fourth leading track from the
$J/\psi \rho^0$ selection in data.  A low-lying tail is observed in
this distribution at large average longitudinal impact parameter
values, indicating simultaneous $J/\psi \rho^0$ production from
separate Pb+Pb interactions.  This distribution is compared with the
same distribution from simulated $\psip \to
\jpsi(\to\ellell)\pi^+\pi^-$ MC events, to check the expected shape
from the single Pb+Pb interaction events: the two
processes clearly arise from the same vertex at a much higher rate
than for separate ones.  A similar process that shares similar topology,
$\gamma\gamma\to\mumu + \rho^0$, has recently been measured by ATLAS~\cite{multi-upc}, highlighting
the importance of these effects.  Further studies will be needed to
assess the relative impact of these processes on the event selection
requirements used by the different experiments.  Notably, the relative
rates for $J/\psi+\ee$ production from a single interaction over
separate interactions can differ substantially from those from $J/\psi
\rho^0$ due to differences between the kinematics of the respective processes.


%

%


\section{Conclusions}
\label{sec:Conclusions}

A first measurement of coherent exclusive $J/\psi$ production in ultraperipheral
collisions in ATLAS is presented.
The measurement uses 79 $\mu$b$^{-1}$ of Pb+Pb collision data recorded at the LHC in 2023 at
$\sqrt{s_{\textrm{NN}}}=5.36$~TeV.  Events were selected with
a special Level 1 trigger only requiring a signal in the
ATLAS transition radiation tracking detector.
Results are corrected for all known backgrounds to the dimuon decay
channel, with the $\jpsi\to\ee$ channel subtracted as a background due
to distortions to the mass peak and reconstruction efficiency from
detector material effects.  Differential cross sections are measured
as a function of $\jpsi$ rapidity in the rapidity interval $|y|<2.5$
and are compared with theory predictions and, after
extrapolation to $\sqrt{s_{\textrm{NN}}}=5.02$~TeV, with previous measurements from other LHC
experiments.
The data are best described by calculations based on the color dipole model.
Significant tension is observed with
ALICE data at mid-rapidity when extrapolating this measurement to
$5.02$~TeV.  This tension may be related to the
effect of additional UPC processes, from the same Pb+Pb interaction, on
the signal exclusivity conditions.

%
%
%


%

%
%
%
%
%
\section*{Acknowledgements}
%

%

%
%

%
%

We thank CERN for the very successful operation of the LHC and its injectors, as well as the support staff at
CERN and at our institutions worldwide without whom ATLAS could not be operated efficiently.

The crucial computing support from all WLCG partners is acknowledged gratefully, in particular from CERN, the ATLAS Tier-1 facilities at TRIUMF/SFU (Canada), NDGF (Denmark, Norway, Sweden), CC-IN2P3 (France), KIT/GridKA (Germany), INFN-CNAF (Italy), NL-T1 (Netherlands), PIC (Spain), RAL (UK) and BNL (USA), the Tier-2 facilities worldwide and large non-WLCG resource providers. Major contributors of computing resources are listed in Ref.~\cite{ATL-SOFT-PUB-2026-001}.

We gratefully acknowledge the support of ANPCyT, Argentina; YerPhI, Armenia; ARC, Australia; BMWFW and FWF, Austria; ANAS, Azerbaijan; CNPq and FAPESP, Brazil; NSERC, NRC and CFI, Canada; CERN; ANID, Chile; CAS, MOST and NSFC, China; Minciencias, Colombia; MEYS CR, Czech Republic; DNRF and DNSRC, Denmark; IN2P3-CNRS and CEA-DRF/IRFU, France; SRNSFG, Georgia; BMFTR, HGF and MPG, Germany; GSRI, Greece; RGC and Hong Kong SAR, China; ICHEP and Academy of Sciences and Humanities, Israel; INFN, Italy; MEXT and JSPS, Japan; CNRST, Morocco; NWO, Netherlands; RCN, Norway; MNiSW, Poland; FCT, Portugal; MNE/IFA, Romania; MSTDI, Serbia; MSSR, Slovakia; ARIS and MVZI, Slovenia; DSI/NRF, South Africa; MICIU/AEI, Spain; SRC and Wallenberg Foundation, Sweden; SERI, SNSF and Cantons of Bern and Geneva, Switzerland; NSTC, Taipei; TENMAK, T\"urkiye; STFC/UKRI, United Kingdom; DOE and NSF, United States of America.

Individual groups and members have received support from BCKDF, CANARIE, CRC and DRAC, Canada; CERN-CZ, FORTE and PRIMUS, Czech Republic; COST, ERC, ERDF, Horizon 2020, ICSC-NextGenerationEU and Marie Sk{\l}odowska-Curie Actions, European Union; Investissements d'Avenir Labex, Investissements d'Avenir Idex and ANR, France; DFG and AvH Foundation, Germany; Herakleitos, Thales and Aristeia programmes co-financed by EU-ESF and the Greek NSRF, Greece; BSF-NSF and MINERVA, Israel; NCN and NAWA, Poland; La Caixa Banking Foundation, CERCA Programme Generalitat de Catalunya and PROMETEO and GenT Programmes Generalitat Valenciana, Spain; G\"{o}ran Gustafssons Stiftelse, Sweden; The Royal Society and Leverhulme Trust, United Kingdom; United States of America.

In addition, individual members wish to acknowledge support from CERN: European Organization for Nuclear Research (CERN DOCT); Chile: Agencia Nacional de Investigaci\'on y Desarrollo (ANID FONDECYT reg. 1230987, FONDECYT 1230812, FONDECYT 1240864, Fondecyt 3240661, Fondecyt Regular 1240721); China: Chinese Ministry of Science and Technology (MOST-2023YFA1605700, MOST-2023YFA1609300), National Natural Science Foundation of China (NSFC - 12175119, NSFC 12275265); Czech Republic: Czech Science Foundation (GACR - 24-11373S), Ministry of Education Youth and Sports (ERC-CZ-LL2327, FORTE CZ.02.01.01/00/22\_008/0004632), PRIMUS Research Programme (PRIMUS/21/SCI/017); EU: H2020 European Research Council (ERC - 101002463); European Union: European Research Council (BARD No. 101116429, ERC - 948254, ERC 101089007), European Regional Development Fund (HE COFUND GA No.101081355, ERDF), European Union, Future Artificial Intelligence Research (FAIR-NextGenerationEU PE00000013), Italian Center for High Performance Computing, Big Data and Quantum Computing (ICSC, NextGenerationEU), Marie Sklodowska-Curie Actions (GAP-101168829); France: Agence Nationale de la Recherche (ANR-21-CE31-0013, ANR-21-CE31-0022, ANR-22-EDIR-0002, ANR-24-CE31-0504-01); Germany: Baden-Württemberg Stiftung (BW Stiftung-Postdoc Eliteprogramme), Deutsche Forschungsgemeinschaft (DFG - 469666862, DFG - CR 312/5-2); China: Research Grants Council (GRF); Italy: Istituto Nazionale di Fisica Nucleare (ICSC, NextGenerationEU), Ministero dell'Università e della Ricerca (NextGenEU 153D23001490006 M4C2.1.1, NextGenEU I53D23000820006 M4C2.1.1, NextGenEU I53D23001490006 M4C2.1.1, SOE2024\_0000023); Japan: Japan Society for the Promotion of Science (JSPS KAKENHI JP22H01227, JSPS KAKENHI JP22H04944, JSPS KAKENHI JP22KK0227, JSPS KAKENHI JP24K23939, JSPS KAKENHI JP24KK0251, JSPS KAKENHI JP25H00650, JSPS KAKENHI JP25H01291, JSPS KAKENHI JP25K01023); Norway: Research Council of Norway (RCN-314472); Poland: Ministry of Science and Higher Education (IDUB AGH, POB8, D4 no 9722), Polish National Science Centre (NCN 2021/42/E/ST2/00350, NCN OPUS 2023/51/B/ST2/02507, NCN OPUS nr 2022/47/B/ST2/03059, NCN UMO-2019/34/E/ST2/00393, UMO-2022/47/O/ST2/00148, UMO-2023/49/B/ST2/04085, UMO-2023/51/B/ST2/00920, UMO-2024/53/N/ST2/00869); Portugal: Foundation for Science and Technology (FCT); Spain: Agencia de Gestión de Ayudas Universitarias y de Investigación (AGAUR - 2023 BP 00141), Generalitat Valenciana (ASFAE/2022/008), Ministry of Science and Innovation (MCIN \& NextGenEU PCI2022-135018-2, MICIN \& FEDER PID2021-125273NB, RYC2019-028510-I, RYC2020-030254-I, RYC2021-031273-I, RYC2022-038164-I), Ministerio de Ciencia, Innovación y Universidades/Agencia Estatal de Investigaci\'on (PID2022-142604OB-C22); Sweden: Carl Trygger Foundation (Carl Trygger Foundation CTS 22:2312), Swedish Research Council (Swedish Research Council 2023-04654, VR 2021-03651, VR 2022-03845, VR 2022-04683, VR 2023-03403, VR 2024-05451), Knut and Alice Wallenberg Foundation (KAW 2018.0458, KAW 2022.0358, KAW 2023.0366); Switzerland: Swiss National Science Foundation (SNSF - PCEFP2\_194658); United Kingdom: Royal Society (NIF-R1-231091); United States of America: U.S. Department of Energy (ECA DE-AC02-76SF00515), Neubauer Family Foundation.

%
%


%
\clearpage
\printbibliography
\clearpage
 
\begin{flushleft}
\hypersetup{urlcolor=black}
{\Large The ATLAS Collaboration}

\bigskip

\AtlasOrcid[0000-0002-6665-4934]{G.~Aad}$^\textrm{\scriptsize 104}$,
\AtlasOrcid[0000-0001-7616-1554]{E.~Aakvaag}$^\textrm{\scriptsize 17}$,
\AtlasOrcid[0000-0002-5888-2734]{B.~Abbott}$^\textrm{\scriptsize 123}$,
\AtlasOrcid[0000-0002-0287-5869]{S.~Abdelhameed}$^\textrm{\scriptsize 119a}$,
\AtlasOrcid[0000-0002-1002-1652]{K.~Abeling}$^\textrm{\scriptsize 55}$,
\AtlasOrcid[0000-0001-5763-2760]{N.J.~Abicht}$^\textrm{\scriptsize 49}$,
\AtlasOrcid[0000-0002-8496-9294]{S.H.~Abidi}$^\textrm{\scriptsize 30}$,
\AtlasOrcid[0009-0003-6578-220X]{M.~Aboelela}$^\textrm{\scriptsize 45}$,
\AtlasOrcid[0000-0002-9987-2292]{A.~Aboulhorma}$^\textrm{\scriptsize 36e}$,
\AtlasOrcid[0000-0001-5329-6640]{H.~Abramowicz}$^\textrm{\scriptsize 157}$
% ...
\bigskip
\\

% ...
$^{17}$Department for Physics and Technology, University of Bergen, Bergen; Norway.\\
% ...
$^{30}$Physics Department, Brookhaven National Laboratory, Upton NY; United States of America.\\
% ...
$^{36}$$^{(a)}$Facult\'e des Sciences Ain Chock, Universit\'e Hassan II de Casablanca;$^{(b)}$Facult\'{e} des Sciences, Universit\'{e} Ibn-Tofail, K\'{e}nitra;$^{(c)}$Facult\'e des Sciences Semlalia, Universit\'e Cadi Ayyad, LPHEA-Marrakech;$^{(d)}$LPMR, Facult\'e des Sciences, Universit\'e Mohamed Premier, Oujda;$^{(e)}$Facult\'e des sciences, Universit\'e Mohammed V, Rabat;$^{(f)}$Institute of Applied Physics, Mohammed VI Polytechnic University, Ben Guerir; Morocco.\\
% ...
$^{45}$Physics Department, Southern Methodist University, Dallas TX; United States of America.\\
% ...
$^{49}$Fakult\"{a}t Physik, Technische Universit{\"a}t Dortmund, Dortmund; Germany.\\
% ...
$^{55}$II. Physikalisches Institut, Georg-August-Universit\"{a}t G\"ottingen, G\"ottingen; Germany.\\
% ...
$^{104}$CPPM, Aix-Marseille Universit\'e, CNRS/IN2P3, Marseille; France.\\
% ...
$^{119}$$^{(a)}$New York University Abu Dhabi, Abu Dhabi;$^{(b)}$United Arab Emirates University, Al Ain; United Arab Emirates.\\
% ...
$^{123}$Homer L. Dodge Department of Physics and Astronomy, University of Oklahoma, Norman OK; United States of America.\\
% ...
$^{157}$Raymond and Beverly Sackler School of Physics and Astronomy, Tel Aviv University, Tel Aviv; Israel.\\
% ...
$^{a}$ Also at Affiliated with an institute formerly covered by a cooperation agreement with CERN.\\
$^{b}$ Also at An-Najah National University, Nablus; Palestine.\\
$^{c}$ Also at Borough of Manhattan Community College, City University of New York, New York NY; United States of America.\\
$^{d}$ Also at Center for Interdisciplinary Research and Innovation (CIRI-AUTH), Thessaloniki; Greece.\\
$^{e}$ Also at Centre of Physics of the Universities of Minho and Porto (CF-UM-UP); Portugal.\\
$^{f}$ Also at CERN, Geneva; Switzerland.\\
% ...

\end{flushleft}


%

%
%

%

%
%
%
%

\end{document}